\documentclass[aps,prd,letterpaper,twocolumn,preprintnumbers,
amsmath,amssymb,nofootinbib,floatfix,showpacs]{revtex4}
\usepackage{graphicx}
\usepackage{amsmath}
\usepackage{tabularx}
\usepackage{longtable}

\newcommand{\mys}[1]{\section{#1}
    \setcounter{equation}{0}}
    \renewcommand{\theequation}{\arabic{section}.\arabic{equation}}
\newcommand{\myappendix}{\appendix
   \renewcommand{\theequation}{\Alph{section}.\arabic{equation}}
   }

\DeclareMathAlphabet   {\mathsc}{OT1}{cmr}{m}{sc}

\def\[{\left [}
\def\]{\right ]}
\def\({\left (}
\def\){\right )}
\newcommand{\lang}{\left\langle}
\newcommand{\rang}{\right\rangle}
\newcommand{\lbr}{\left\{}
\newcommand{\rbr}{\right\}}
\newcommand{\oline}[1]{\overline{#1}}

\newcommand{\wtd}[1]{\widetilde{#1}}

\newcommand{\wh}[1]{\widehat{#1}}

\newcommand{\GeV}      {~\mathrm{GeV}}

\newcommand{\SM}       {\mathsc{sm}}

\newcommand{\PL}       {\mathsc{pl}}

\newcommand{\STR}      {\mathsc{str}}

\newcommand{\order}{\mathcal{O}}

\newcommand{\gappeq}{\mathrel{\rlap {\raise.5ex\hbox{$>$}}
{\lower.5ex\hbox{$\sim$}}}}
\newcommand{\lappeq}{\mathrel{\rlap{\raise.5ex\hbox{$<$}}
{\lower.5ex\hbox{$\sim$}}}}

\newlength{\dummysp}
\newcommand{\tr}{\mathop{{\hbox{Tr} \, }}\nolimits}

\newcommand{\stxt}[1]{\mathop{\hbox{{\scriptsize #1}}}\nolimits}
\newcommand{\bbar}[1]{{\overline{#1}}}
\newcommand{\half}{\frac{1}{2}}

\newcommand{\beq}{\begin{eqnarray}}
\newcommand{\eeq}{\end{eqnarray}}
\newcommand{\nnn}{ \nonumber \\ }
\newcommand{\p}{{\partial}}

\newcommand{\Zbf}{{{\bf Z}}}

\newcommand{\e}{{\epsilon}}

\newcommand{\vev}[1]{{\langle #1 \rangle}}

\newcommand{\ord}[1]{{{\cal O}(#1)}}
\newcommand{\myref}[1]{(\ref{#1})}

\newcommand{\bfe}[1]{\vspace{5pt} {\bf #1 \hspace{2pt}}}

\newcommand{\ben}{\begin{enumerate}}
\newcommand{\een}{\end{enumerate}}

\newcommand{\bit}{\begin{itemize}}
\newcommand{\eit}{\end{itemize}}

\newcommand{\pX}{\p X}
\newcommand{\pXb}{\bbar{\pX}}
\newcommand{\eV}      {~\mathrm{eV}}

\begin{document}
\preprint{MCTP-05-01, UPR-1106T}

\title{Massive neutrinos and (heterotic) string theory 
}

\author{Joel Giedt}
\affiliation{Department of Physics, University of Toronto, 60 St.
George St., Toronto, ON M5S 1A7, Canada}

\author{G.~L.~Kane}
\affiliation{Michigan Center for Theoretical Physics, University
of Michigan, Ann Arbor, MI 48109, USA}

\author{Paul~Langacker}
\author{Brent D.~Nelson} 
\affiliation{Department of Physics \& Astronomy, University of
Pennsylvania, Philadelphia, PA 19103, USA 
}

\date{\today 
}
\begin{abstract}
String theories in principle address the origin and values of the
quark and lepton masses. Perhaps the small values of neutrino
masses could be explained generically in string theory even if it
is more difficult to calculate individual values, or perhaps some
string constructions could be favored by generating small neutrino
masses. We examine this issue in the context of the well-known
three-family standard-like $Z_3$ heterotic orbifolds, where the
theory is well enough known to construct the corresponding
operators allowed by string selection rules, and analyze the
D-~and F-flatness conditions. Surprisingly, we find that a simple
see-saw mechanism does not arise. It is not clear whether this is
a property of this construction, or of orbifolds more generally,
or of string theory itself. Extended see-saw mechanisms may be
allowed; more analysis will be needed to settle that issue. We
briefly speculate on their form if allowed and on the possibility
of alternatives, such as small Dirac masses and triplet see-saws.
The smallness of neutrino masses may be a powerful probe of string
constructions in general. We also find further evidence that there
are only~20 inequivalent models in this class, which affects the
counting of string vacua.
\end{abstract}


\maketitle


\mys{Introduction}
String theory proposes to provide a well-defined underlying theory
for elementary particle physics. As such it is obligated to
provide an understanding for the phenomena we see at accessible
energy scales, including the origin of fermion masses and mixings.
In particular, one should be able to identify the mechanism that
explains the smallness of neutrino masses as a natural outcome in
some class of explicit string constructions.  In this paper we
perform a study of a particular class of real string constructions
in a top-down manner, and search for the couplings necessary to
generate the ``minimal see-saw'' mechanism for neutrino masses (to
be defined more precisely below). Though we are mindful that this
is not the only possible method of achieving very light neutrinos,
it does lead naturally to very small masses (though not
necessarily to large mixing angles) and it is the basis of the
vast majority of phenomenological studies of neutrinos in the
literature.\footnote{For recent reviews of the neutrino
oscillation data and models of their masses and mixings,
see~\cite{Fisher:1999fb,Murayama:2002bj,Gonzalez-Garcia:2002dz,King:2003jb,Mohapatra:2004vr}.}
The minimal see-saw requires a well-defined set of fields and
couplings to be present in the low-energy theory. In particular,
it requires the simultaneous presence of both Dirac mass terms and
large Majorana mass terms for the right-handed neutrinos. There is
no Standard Model symmetry to forbid such couplings. Large
Majorana masses might, however, be forbidden by extensions of the
low energy theory, such as an additional $U(1)'$ gauge
symmetry~\cite{Kang:2004ix}.
%
Their possible existence forms a useful probe of the much more
restrictive string constructions.

Sadly, string theory has been largely silent on the issue of
neutrino mass since the subject was first raised in the context of
heterotic strings nearly twenty years
ago~\cite{Witten:1985bz,Font:1989aj}. The reason for this silence
is not hard to understand: the issue of flavor is perhaps the most
difficult phenomenological problem to study in explicit, top-down
string constructions -- and neutrino masses are just one aspect of
this problem. To begin such a study requires that many things be
worked out: one needs not just the spectrum of massless states,
but also their charges under all Abelian symmetries (properly
redefined so that only one linear combination of $U(1)$ factors is
anomalous). To obtain the superpotential couplings to very high
order the string selection rules for the particular construction
must be worked out and put into a form amenable to automation.
Obtaining these working ingredients takes time and effort, though
the techniques are well known. Certain parts of this process have
been completed and discussed in the literature for several string
models. The most comprehensive study of weakly-coupled heterotic
models with semi-realistic gauge groups and particle content are
the free-fermionic constructions (see for
example~\cite{Cleaver:1998gc,Cleaver:1998sm,Cleaver:1998sa,Cleaver:2000aa}
and references therein) and the bosonic standard-like $Z_3$
orbifold constructions (see for
example~\cite{Font:1989aj,Giedt:2001zw} and references therein).
In particular, a systematic study of the spectra in the
phenomenologically promising BSL$_A$ class of the $Z_3$ orbifold
has been performed by one of the authors of this paper. Thus we
have at our disposal these results and here we will exploit them
to perform a systematic study of the superpotential couplings and
flat directions\footnote{These flat directions are only
approximate ({\em i.e.} through degree~9); see
Appendix~\ref{violations}.} -- with particular emphasis on the
issue of neutrino mass. We will define this BSL$_A$ class more
properly in Section~\ref{bsladef}.

But merely working out the allowed superpotential couplings
(itself a tedious task) is not sufficient for studying neutrino
masses. The minimal see-saw calls for a very special type of
coupling: a supersymmetric bilinear Majorana mass term. Such terms
do not arise from string theory for the states in the massless
spectrum. Thus it must be that this term arises {\em dynamically}
through the vacuum expectation value (vev) of some field or
fields. This means we must consider the issue of flat directions
in the space of chiral matter fields. To be more precise, in
semi-realistic string constructions we are inevitably faced with
an enormous vacuum degeneracy that we do not know how to resolve
from first principles.  So even once we have
\begin{itemize}
\item[(i)]
assumed a particular string construction,
\item[(ii)]
assumed a particular compactification,
and
\item[(iii)]
assumed (or better yet, determined) the background values for
string moduli,
\end{itemize}
we are still
\begin{itemize}
\item[(iv)] faced with a wealth of D- and F-flat directions in the space
of chiral matter fields.
\end{itemize}
These flat directions are combinations of background field values
for which the classical scalar potential vanishes and
supersymmetry is maintained. There are typically many such
directions, all degenerate and all consistent vacuum
configurations of the string construction. By setting certain
chiral superfields to background values that do {\em not} lie
along such a flat direction, one is attempting to expand about an
inappropriate configuration -- in particular, a configuration in
which supersymmetry is spontaneously broken at a high scale (not
to mention a configuration which is not a valid point for a
saddle-point expansion). A minimum of the classical potential, and
thus a classical vacuum, will be the supersymmetric
one.\footnote{Here we assume that there is a supersymmetric
minimum.  Also, in the presence of low-scale
supersymmetry-breaking effects, the minimum of the effective
potential may be shifted slightly away from a supersymmetric
minimum.} Therefore, to the extent that superpotential couplings
of the MSSM arise from terms involving one or more vevs, they must
occur along such flat directions, and the issue of fermion masses
and flavor is intricately tied to the issue of vacuum selection in
string theory. Hence the great difficulty in studying the issue of
neutrino masses.

Only a handful of investigations into neutrino masses in {\em
explicit} top-down string constructions have been performed,
though there are many more examples of ``string inspired''
bottom-up studies, and it is worthwhile to review these instances
before proceeding. Two noteworthy examples in intersecting brane
constructions are that of Iba\~nez et al.~\cite{Ibanez:2001nd} and
that of Antoniadis et al.~\cite{Antoniadis:2002qm}. These are both
nonsupersymmetric constructions with low string scales. In the
first case Majorana neutrino masses are forbidden by a residual
global symmetry broken only by chiral symmetry breaking effects so
that masses can only be of the Dirac type. In the second case
Majorana couplings are again forbidden, but a large internal
dimension is used to justify the smallness of neutrino masses.
Heterotic examples come closer to realizing the standard see-saw
paradigm. The most complete top-down analyses involve
free-fermionic constructions of Ellis et
al.~\cite{Ellis:1997ni,Ellis:1998nk} and Faraggi et
al.~\cite{Faraggi:1990it,Faraggi:1993zh,Coriano:2003ui} in which a
detailed treatment of flat directions was performed. In both cases
some assumptions about strong dynamics in the hidden sector need
to be made in order to populate the neutrino mass matrix. In
addition, the latter set of models involves an extended set of
fields that are not of the minimal see-saw variety. In both of
these heterotic cases (as well as the recent heterotic
construction of Kobayashi et al.~\cite{Kobayashi:2004ud}) several
right-handed neutrino species are involved, where by ``different
species'' we mean that right-handed neutrinos with different
gauge-charges with respect to some extension of the Standard Model
gauge group are involved. The point to be made here is not to say
that these are impossibilities or that these examples cannot
explain the smallness of neutrino masses. It is rather to
emphasize that in the (very few) extant string examples neutrino
and lepton mass matrices arise that look very different from those
expected from a typical GUT ansatz, and in some cases different
from those stemming from a minimal see-saw ansatz.

Our focus will be very much different. In the heterotic-based
papers mentioned above one or two flat directions were chosen for
further study on the basis of certain phenomenological virtues
that that are not directly concerned with neutrino masses, such as
the projecting of certain exotic matter states from the spectrum
or the desire to accommodate realistic quark masses. Neutrino
masses are an afterthought, and it is not clear whether the lack
of a simple see-saw is indicative of the string construction or
simply the flat direction chosen. Here we will take the uniqueness
of the neutrino self-coupling of the minimal see-saw as a guide
and study a very large class of flat-directions, in a large set of
models, in search of precisely this coupling. By putting the issue
of neutrino masses as the primary consideration we are thus
examining the question of whether it is possible to infer the
existence of small neutrino masses as a reasonable outcome of a
consistent, explicit string construction independent of other
questions of low energy phenomenology. (Of course, our preference
and ultimate goal would be to find a theory in which the
conspiracy of operators, couplings, charges, etc., was such that
neutrinos almost ``had'' to be light, rather than it being an
accident.) If the answer is positive then it would also be of
interest to study whether the string constraints provide any
insight into such issues as the existence of two large mixing
angles, the nature of the mass hierarchy (ordinary or inverted or
approximately degenerate) and the relation (if any) to the quark
and charged lepton masses and mixings.

We now summarize the content and results of this article. In
Section~\ref{bsladef} we define and motivate the class of
heterotic orbifold models that we analyze and explain the
remarkable fact that from a starting point of thousands of
possibilities one is left to consider only twenty inequivalent
cases. A definition of the minimal see-saw and a description of
the algorithm we use for finding it is described in
Section~\ref{sca}, where we also provide various details regarding
flat directions and couplings in the models. Two cases of the
twenty allowed for the possibility of a Majorana coupling along a
flat direction, though both ultimately fail to provide the minimal
see-saw or realistic neutrino masses for a variety of reasons.
Nevertheless we investigate both in some detail and describe their
successes and ultimate failures in Section~\ref{cases}. We comment
on the possible implications and alternatives in the concluding
Section~\ref{conc}. Supporting material is provided in three
appendices.  In Appendix~\ref{tables} we list the relevant fields
in the spectra for representative models of the two promising
cases of Section~\ref{cases}. In Appendix~\ref{violations} we
discuss the extent to which the flat directions we identify should
be considered approximately flat. Finally, in
Appendix~\ref{selapp} we provide a brief and accessible review of
the string selection rules as they apply to superpotential
couplings in the effective supergravity. We show that it is
possible to reduce these (conveniently) to gauge invariance and a
set of triality (i.e., $Z_3$) invariances.

\mys{BSL$_A$ models of the $Z_3$ orbifold} \label{bsladef}
As mentioned in the introduction, to systematically study the
issue of flavor in general -- and neutrino masses in particular --
requires a thorough knowledge of many aspects of the low-energy
(4D) theory. To date one of the few classes of string
constructions where this process has been systematically performed
are standard-like models obtained from $Z_3$
orbifolds~\cite{Dixon:1985jw,Dixon:1986jc} of the $E_8 \times E_8$
heterotic string~\cite{Gross:1984dd}. The construction is {\it
bosonic} because of the way in which fields in the underlying 2D
conformal field theory are realized, and {\it symmetric} because
left-moving and right-moving 2D conformal field theory degrees of
freedom associated with the compact 6D space are treated
symmetrically. There is an Abelian embedding of the orbifold
action (space group) into the gauge degrees of freedom through a
shift embedding $V$ with two discrete Wilson lines $a_1$ and
$a_3$~\cite{Ibanez:1986tp,Ibanez:1987pj}. Note that there are only
three independent Wilson lines in the $Z_3$ orbifold. Because the
third Wilson line is set to zero, one automatically obtains a
three generation model -- part of the reason for the
phenomenological interest in the $Z_3$ construction.

The twist vector which represents the orbifold action and the
Wilson lines are embedded in the $E_8 \times E_8$ root torus, and
the result is a breaking of this group to a product of subgroups.
The surviving groups emerge according to $E_8^{(1)} \to G_{\rm
obs}$ and $E_8^{(2)} \to G_{\rm hid}$. There are a vast number of
such consistent embeddings, but it has been shown in the $Z_3$
case that many of them are actually
equivalent~\cite{Casas:1989wu}. The bosonic standard-like models
of type~A (BSL$_A$ models -- a designation coined
in~\cite{Giedt:2001zw}) are those models of this class which have
$G_{\rm obs} = SU(3) \times SU(2) \times U(1)^5$ and (three
generations of) $(3,2)$ representations in order to accommodate
the quark doublets. As it turns out, with the choice of $G_{\rm
obs}$ that has been imposed, the $(3,2)$ representations always
occur in the untwisted sector~\cite{Kim:1992en}.

Three generation models of this type have appeared extensively in
the literature on semi-realistic heterotic
orbifolds~\cite{Ibanez:1987sn,Font:1988mm,Casas:1988se,Casas:1988hb,Font:1989aj}.
A complete enumeration of consistent embeddings into the first
$E_8$ was given in~\cite{Casas:1989wu}. However, it is in general
not possible to place all the fields of the MSSM matter content
exclusively in the untwisted sector. Thus it is necessary to work
out the spectrum of twisted sector states as well, and to
therefore provide all possible completions of the embeddings
in~\cite{Casas:1989wu} to the hidden sector $E_8$ factor as
well.\footnote{When a nontrivial embedding is chosen, the
low-energy gauge theory results from a twisted affine Lie algebra
in the underlying 2D conformal field theory. In this case the
weights of the states under the original $E_8 \times E_8$ Cartan
elements are shifted by fractional amounts~$0 \mod 1/3$.  The
twisted sector states are then typically charged under both $E_8$
factors, so that it is necessary to know the ``hidden'' sector
embedding to obtain the states that are also charged under the
``observable'' sector Cartan elements; i.e., one loses the clear
distinction between hidden and observable sector states.} A
complete enumeration of consistent completions of these embeddings
into the second $E_8$ was given in~\cite{Giedt:2000bi}. There it
was found that only five possibilities for $G_{\rm hid}$ exist.
For one of the five possibilities, the non-abelian part of $G_{\rm
hid}$ only contains $SU(2)$ factors. We do not regard this as a
viable hidden sector for dynamical supersymmetry breaking by
gaugino condensation, since the condensation scale will be far too
low to provide a reasonable scale of supersymmetry
breaking~\cite{Gaillard:1999et}. Therefore only the four remaining
possibilities for $G_{\rm hid}$ are of interest to us. With this
restriction, it was found that there are just~175 models.

A systematic study of several properties of these models was given
in~\cite{Giedt:2001zw}. In particular the complete massless
spectrum of chiral matter was determined for each of the~175
models, where it was found that only 20~distinct sets of
representations occur for the 175~models. Furthermore, the 4D
theories generated by the different embeddings in each of the
twenty classes had several identical physical properties. This was
interpreted as an indication that models in the same pattern are
actually equivalent. We will refer to these 20~cases as {\em
patterns} of the BSL$_A$ master class.

\begin{table}[tb]
\caption{\label{tbl:bsla} \textbf{Summary of BSL$_A$ patterns.}
For each pattern we give the number of models in the pattern, the
hidden sector gauge group, the scale of the anomalous $U(1)$
FI-term and the number of distinct species of chiral matter
superfields.}
\begin{tabular}{|c|cccc|} \hline
\parbox{1cm}{Pattern} & \parbox{1cm}{No.} &
\parbox{1.5cm}{$G_{\rm hid}$} & \parbox{1.2cm}{$r_{\rm FI}$} &
\parbox{1cm}{Species} \\ \hline \hline
1.1 & 7 & $SO(10)\times U(1)^3$ & No $U(1)_X$ & 51 \\
1.2 & 7 & $SO(10)\times U(1)^3$ & 0.15 & 76 \\ \hline
2.1 & 10 & $SU(5)\times SU(2)\times U(1)^3$ & 0.09 & 64 \\
2.2 & 10 & $SU(5)\times SU(2)\times U(1)^3$ & 0.10 & 66 \\
2.3 & 7 & $SU(5)\times SU(2)\times U(1)^3$ & 0.10 & 65 \\
2.4 & 7 & $SU(5)\times SU(2)\times U(1)^3$ & 0.13 & 60 \\
2.5 & 6 & $SU(5)\times SU(2)\times U(1)^3$ & 0.14 & 61 \\
2.6 & 6 & $SU(5)\times SU(2)\times U(1)^3$ & 0.12 & 51 \\ \hline
3.1 & 12 & $SU(4)\times SU(2)^2\times U(1)^3$ & 0.07 & 58 \\
3.2 & 5 & $SU(4)\times SU(2)^2\times U(1)^3$ & 0.12 & 57 \\
3.3 & 10 & $SU(4)\times SU(2)^2\times U(1)^3$ & 0.12 & 57 \\
3.4 & 5 & $SU(4)\times SU(2)^2\times U(1)^3$ & 0.13 & 53 \\ \hline
4.1 & 7 & $SU(3)\times SU(2)^2\times U(1)^4$ & 0.10 & 61 \\
4.2 & 12 & $SU(3)\times SU(2)^2\times U(1)^4$ & 0.09 & 62 \\
4.3 & 7 & $SU(3)\times SU(2)^2\times U(1)^4$ & 0.07 & 63 \\
4.4 & 15 & $SU(3)\times SU(2)^2\times U(1)^4$ & 0.12 & 59 \\
4.5 & 17 & $SU(3)\times SU(2)^2\times U(1)^4$ & 0.11 & 61 \\
4.6 & 13 & $SU(3)\times SU(2)^2\times U(1)^4$ & 0.12 & 60 \\
4.7 & 6 & $SU(3)\times SU(2)^2\times U(1)^4$ & 0.11 & 62 \\
4.8 & 6 & $SU(3)\times SU(2)^2\times U(1)^4$ & 0.12 & 53 \\ \hline
\end{tabular}
\end{table}

We give a summary of these twenty patterns in
Table~\ref{tbl:bsla}, organized by the hidden sector gauge group
$G_{\rm hid}$, with the number of different embeddings in each
pattern given in the first column. In each of the~175 individual
embeddings the $U(1)$ charges of the spectrum can be obtained and
the anomaly isolated to one Abelian factor $U(1)_X$. This anomaly
is cancelled~\cite{Dine:1987xk,Dine:1987gj,Atick:1987gy} by the
Green-Schwarz (GS) mechanism~\cite{Green:1984sg}, which involves a
Fayet-Iliopoulos (FI) term in the 4D Lagrangian. This term can be
calculated from the known spectrum and is given by
\begin{equation}
\xi_{\rm FI} = \frac{g_{\STR}^2 \tr Q_X}{192\pi^2} M_{\PL}^2 ,
\label{FIterm} \end{equation}
where $g_{\STR}$ is the (unified) string coupling just below the
compactification scale and $M_{\PL}$ is the reduced Planck mass.
We will discuss the relevance of this particular mass scale in
Section~\ref{cases}. The value of $\xi_{\rm FI}$ depends on the
vacuum expectation value of the dilaton, which provides the
determination of $g_{\STR}$. If we take the unification of
couplings in the MSSM as a rough guide to what this vev might be,
we can make the approximation $g_{\STR}^2 \simeq 0.5$. Then the
value of the ratio $r_{\rm FI} = \sqrt{|\xi_{\rm FI}|}/M_{\PL}$
can be calculated; we provide the numerical value of this factor
in the third column of Table~\ref{tbl:bsla}.  In the final column
we give the total number of different species of chiral superfield
in each of the individual models for the pattern. Note that the
total number of fields would then involve a three-fold replication
of these species (except that for twisted oscillator states there
is a 9-fold multiplicity).

When the gauge anomaly is isolated to a single Abelian factor, the
only nominal difference between different members of each pattern
is the apparent charges under the various $U(1)$ factors in
$G_{\rm obs} \times G_{\rm hid}$. Yet the value of $\tr Q_X$ is
identical for each member of a given pattern, suggesting that a
basis exists for which these charges -- and perhaps those of all
the Abelian factors -- would, in fact, be identical. In this case
the members of each pattern would truly be redundant models; in
this work we will find further evidence for this conjecture.

Clearly, then, the bosonic standard-like models of the $Z_3$
orbifold are an ideal starting point for a dedicated study of
neutrino masses: they already contain the Standard Model particle
content and gauge group (though with much more besides), the vast
number of possibilities has been reduced to a tractable number
through much past research, and many of the key ingredients needed
for our analysis are already known. Indeed, these properties have
made this class a laboratory for other recent work in string
phenomenology~\cite{Giedt:2000es,Munoz:2001yj,Abel:2002ih,Munoz:2003au,Forste:2004ie}.

\mys{The search for neutrino mass couplings} \label{sca}
We now know with certainty that some neutrinos have mass and that
the different flavors of neutrinos mix with one another with large
mixing angles. We further know that the differences in the squared
masses of the physical eigenstates are extraordinarily small: on
the order of $10^{-3} \eV^2$ for the mass difference that explains
the atmospheric oscillation data and $10^{-5} \eV^2$ for the mass
difference that explains the solar neutrino oscillation data. One
possible explanation is that the masses themselves are of this
order, and indeed cosmological observations of large scale
structure constrain the sum of the physical masses to be on the
order of a few $\times\; 0.1 \eV$~\cite{Seljak:2004xh}. The
fantastic smallness of these numbers, in comparison to the masses
of the quarks and charged leptons, seems to call for an
explanation dramatically different from those of other Standard
Model fields.

\subsection{The minimal see-saw}
\label{seesaw}

While it is logically possible that neutrinos get their masses
solely through electroweak symmetry breaking, with extremely small
Yukawa couplings to Higgs states and right-handed (Dirac)
neutrinos, the preferred explanation has long been the see-saw
mechanism~\cite{GRS,Yanagida}. In this scenario one assumes the
existence of heavy (Majorana) neutrinos which are singlets under
the Standard Model gauge group $G_{\SM}$ which play the role of
right-handed neutrinos. If these heavy states have $\order(1)$
Yukawa couplings to the lepton and Higgs doublets, then
integrating them out of the effective theory produces sufficiently
small effective neutrino masses for the light states. In a
supersymmetric context we can cast this as an effective neutrino
mass superpotential which takes the form
\begin{equation}
W_{\stxt{eff}} = ( \nu_i, N_i) \begin{pmatrix} 0 & (m_D)_{ij} \cr
(m_D)_{ji} & (m_M)_{ij} \cr \end{pmatrix} \binom{ \nu_j}{ N_j }
\label{simss} \end{equation}
where one assumes $m_D \ll m_M$ in order to produce the desired
light eigenvalues. Here, the $\nu_i$ are the neutrino superfields
associated with the Minimal Supersymmetric Standard Model (MSSM),
and the $N_i$ are (charge-conjugated) right-handed neutrino
superfields. One typically assumes that $\nu_i$ runs over three
generations of fields and that there are three (or possibly mode)
$N_j$. We will assume three generations of each type of field,
given the three-generation construction that we appeal to. We will
refer to the neutrino system defined by these assumptions and the
matrix~(\ref{simss}) as the ``minimal see-saw.'' The overwhelming
majority of the vast literature on neutrino phenomenology is based
on this minimal paradigm. We wish to study whether it is possible
to embed this scenario in a string-derived model -- say, a BSL$_A$
model.

Once we introduce string theory, we are confronted with a number
of chiral superfields beyond the states of the MSSM. There are
many potential candidates for the right-handed neutrino fields
$N_i$. In fact, typically half the species in the models of
Table~\ref{tbl:bsla} are singlets of the Standard Model gauge
group -- though none of them are singlets under {\em all} of the
Abelian gauge factors simultaneously. In this they are distinct
from the various moduli of the string theory. These states are
also represented by chiral superfields and are singlets under all
gauge symmetries. Could these be candidates for right-handed
neutrinos?

While not a logical impossibility, we argue that a viable model of
neutrino mass is unlikely to involve these fields. Moduli fields
have no superpotential couplings at the perturbative level, so the
types of Yukawa interactions that can give rise to the
matrix~(\ref{simss}) are absent at this level. Furthermore, the
string moduli are likely to receive a mass only after
supersymmetry is broken, and thus we might expect typical values
in the matrix $(m_M)_{ij}$ to be $\order$(TeV). The entries in the
matrix $(m_D)_{ij}$ would then need to be extremely small to
explain the observed neutrino mass differences. Thus we will
search for the needed couplings among the fields that have been
summarized in Table~\ref{tbl:bsla}.

As mentioned above, bare mass terms $W = m_M \Phi \Phi$ with $m_M
\ll M_{\PL}$ do not arise in a natural way from the underlying
string theory. Thus our first task is to identify a degree $n \geq
3$ coupling that would yield an effective Majorana mass term
\begin{equation}
\vev{S_1 \cdots S_{n-2}} N N , \label{hekr}
\end{equation}
where we have suppressed the generation labels associated with the
3-fold degeneracy of the spectrum. The principal questions that we
address in this section are:
\begin{itemize}
\item[(1)]
Is it possible to get the simplest sort of Majorana mass
couplings~(\ref{hekr}) in the BSL$_A$ models?
\item[(2)]
Since vevs of $G_{\SM}$ singlets $\vev{S_i}$ are necessary, we
must simultaneously ask:  Are these vevs consistent with D- and
F-flatness?
\end{itemize}

\subsection{Flat direction scan and analysis}
\label{flat}

To obtain answers to questions~(1) and~(2), we have studied in
detail all allowed superpotential couplings and an elementary
class of flat directions up to a certain order (described below)
for a representative sample (3 models from each of the~20
representation patterns) of the~175 models in the BSL$_A$ class.
Though straightforward, the computation is very tedious and
impossible without automation; it took weeks for the C/C++
routines to run on a Pentium~4 processor. Some idea of the scale
of the project will be evident in the discussion below, since a
fringe benefit of the analysis is a count of couplings and flat
directions for each model in which interesting aspects of the
BSL$_A$ models emerge. It should also be stated that none of the
analysis made here requires a detailed knowledge of the strength
of couplings.  For a study of class of flat directions that we
consider, it is sufficient to know the selection rules.

%
As is well known, D-flat directions are easily and completely
classified by analytic
invariants~\cite{Buccella:1982nx,Gatto:1986bt,Procesi:hr,Luty:1995sd}.
To each holomorphic gauge invariant $I(\Phi)$ of the chiral
superfields $\Phi_1, \ldots, \Phi_n$ in the theory corresponds a
D-flat direction, given by
\begin{equation} \vev{K_i} = c \vev{I_i}
\end{equation}
where $c$ is a universal constant, $K_i = \p K / \p \Phi_i$
with $K$ the K\"ahler potential, and $I_i = \p I / \p \Phi_i$. Of
course, $c$ can be absorbed into the definition of $I$. It is an
undetermined parameter, whose magnitude corresponds to the scale
of the breaking.  Energetically, all scales are equally favored.

In the case where there is an anomalous $U(1)_X$, a slight
modification is required.  We choose a basis of $U(1)$ charges
where only one, $Q_X$, is anomalous and $\tr Q_X > 0$. We express
the invariant $I$ as a sum of monomials $I^{(A)}$ in the fields
\beq I = \sum_A c_A I^{(A)}. \eeq D-flatness is satisfied if and
only if: (i)~each $I^{(A)}$ is gauge invariant with respect to all
non-anomalous factors of the gauge group, and (ii)~at least one of
the $I^{(A)}$ has strictly negative $Q_X$ charge. The vanishing of
the $Q_X$ D-term imposes one real constraint on the $c_{A}$; an
overall phase among the $c_{A}$ can be removed by going to
unitarity gauge with respect to the $U(1)_X$. The remaining
degeneracy of solutions corresponds to flat directions, termed
elsewhere as {\it D-moduli}~\cite{Gaillard:2000na,Giedt:2002jp}.

In our analysis of effective Majorana neutrino mass couplings, we
restrict our attention to the case where $I$ is a single monomial
satisfying this invariant condition, with $Q_X(I) <0$ if an
anomalous $U(1)$ exists. We refer to this product of fields as an
{\it I-monomial}. Thus we examine only special points in the
D-moduli space. Polynomials that are linear combinations of the
I-monomials correspond to a more general class of D-flat
directions~\cite{Cleaver:1997jb,Cleaver:1997nj}. Since these
generalizations allow for more fields to get vevs, they might
provide new paths to obtain~(\ref{hekr}). However, this more
complicated scenario involves significantly more analysis. While
it is a sensible follow-up to the present study, for practical
reasons we leave it to future work. To simplify our analysis, we
impose {\it stringent F-flatness}~\cite{Cleaver:2000aa}. That is
to say, we do not permit the vev of any monomial in $\p W/\p
\phi_i$ to contribute a nonzero term (to the order we study). This
sufficient but not necessary condition is a further restriction to
special points in D-moduli space, which is nevertheless a class
with a large number of elements.

In our automated search, we adopted the following procedure:
\begin{description}
\item[Step 1] We generated a complete list of I-monomials that (i)
contained only fields neutral under~$SU(3) \times SU(2) \in G_{\rm
obs}$ and, (ii) had degree less than or equal to ten.
\item[Step 2]  All superpotential couplings allowed by
selection rules (see Appendix~\ref{selapp}) were generated, up to
and including degree nine.
\item[Step 3] We eliminated from the list of I-monomials all those
that would violate stringent F-flatness with respect to the
superpotential couplings generated in Step~2.  The remaining
I-monomials specify our list of D- and F-flat directions.
\item[Step 4] For each flat direction that survived Step~3, we
searched for couplings from the list generated in Step~2 that
would provide an effective Majorana mass coupling of the
form~(\ref{hekr}), where the vevs $\vev{S_i}$ were each contained
in the I-monomial of the given flat direction. The repeated field
in the coupling then becomes a candidate right-handed neutrino
$N$.
\item[Step 5] If the candidate $N$ fields were not singlets of the
non-abelian factors of $G_{\rm hid}$, we also checked that the
flat direction broke those factors of the gauge group so that $N$
could be an effective gauge singlet along the flat direction.
\end{description}

\begin{table}[tb]
\caption{\label{spcpn} \textbf{Number of allowed superpotential
couplings by degree.} For each pattern of Table~\ref{tbl:bsla} we
give the number of superpotential coupling at leading order
(degree three) through degree nine allowed by the string selection
rules (note that there were no degree~five couplings allowed for
any pattern).}
\begin{tabular}{|c||c|c|c|c|c|c|} \hline
\parbox{1cm}{Pattern} & \parbox{0.8cm}{3} & \parbox{0.8cm}{4} & \parbox{1cm}{6} &
\parbox{1cm}{7} & \parbox{1cm}{8} & \parbox{1.3cm}{9} \\ \hline \hline
1.1 & 113 & 24 & 21329     & 23768  & 1697 & 3380308 \\
1.2 & 97 & 12 & 13968 & 4418  & 498   & 1552812 \\ \hline
2.1 & 67 & 10 & 5188 & 3515 & 162   & 342186 \\
2.2 & 80 & 11 & 7573 & 3066 & 272   & 582326 \\
2.3 & 75 & 10 & 6508 & 2874 & 250 & 467020 \\
2.4 & 53    & 0     & 2795 & 360 & 0 & 119454 \\
2.5 & 58 & 6     & 3363  & 688 & 26 & 150838 \\
2.6 & 31 & 0     & 642 & 0 & 0 & 10976 \\ \hline
3.1 & 54    & 4 & 2749 & 768   & 21 & 119973 \\
3.2 & 43 & 2     & 1758  & 291       & 9 & 59182 \\
3.3 & 48 & 4 & 2187  & 393 & 20    & 81497 \\
3.4 & 31    & 8     & 750       & 375 & 42    & 15074 \\ \hline
4.1 & 50 & 3 & 2090 & 693       & 14    & 81222 \\
4.2 & 62 & 6 & 3206 & 793 & 38    & 143257 \\
4.3 & 55 & 5     & 2516  & 613 & 15 & 100793 \\
4.4 & 38    & 2 & 1137  & 147 & 3 & 28788 \\
4.5 & 48 & 0 & 1872  & 0 & 0 & 62597 \\
4.6         & 47 & 0 & 1738 & 50 & 0 & 51970 \\
4.7 & 53 & 0 & 2219 & 0 & 0 & 76244 \\
4.8         & 21 & 0     & 301 & 0         & 0 & 4120 \\ \hline
\end{tabular}
\end{table}

The result of this procedure was a success for only~2 of the~20
patterns in these models: Pattern~1.1 and~2.6. This is already a
remarkable result. But this is not sufficient to claim that the
minimal see-saw has been discovered. In cases where we find in the
affirmative on the two questions just posed at the end of
Section~\ref{seesaw}, other questions remain:
\begin{itemize}
\item[(3)]
The vevs required for \myref{hekr}, and any others that are
required for flatness, generally break some of the $U(1)$ factors
in the model.  A question connected with this is: Does a $U(1)$
survive that will serve as electroweak hypercharge $U(1)_Y$, and
in particular is $N$ a singlet under this group?
\item[(4)]
Does the candidate Majorana neutrino $N$ also have the requisite
$H_u L N$ Dirac couplings to $SU(2)$ doublets so as to produce the
$m_D$ entries in \myref{simss}?
\item[(5)]
Do the remainder of the Standard Model particles have the proper
charge assignments under the candidate $U(1)_Y$?
\end{itemize}

If questions (3)-(5) can be answered affirmatively, then we have
the minimal see-saw. Note that we are not demanding anything about
the remaining Yukawa couplings of the MSSM. Of course a truly
realistic model must not only possess such superpotential terms,
it must also possess them in such a way as to give rise to the
observed hierarchies between quark, charged lepton and neutral
lepton masses, and the observed large (small) leptonic (quark)
mixings. As mentioned in the introduction, we are here making
neutrinos our principal focus.

The two patterns that were successful in answering questions~(1)
and~(2) will be discussed in more detail in Section~\ref{cases}.
Here we wish to remark on a few aspects of the flat direction
analysis that deserve some comment. Our procedure clearly produced
a wealth of data on couplings and flat directions for all of the
models in the BSL$_A$ class. A sense of the size of the project
can be seen in the number of superpotential couplings, allowed by
all selection rules, that needed to be studied. These are the
results of Step~2 above, and are listed in Table~\ref{spcpn}. Many
of the higher order couplings are just products of lower order
invariants. However, even taking this into account, the number of
invariants is impressive. That Patterns~2.6, 4.5, 4.7 and~4.8 only
have superpotential couplings whose degree is a multiple of three
follows directly from the string selection rules, as discussed in
Appendix~\ref{selapp}. It is unclear to us why Patterns~2.4
and~4.6 lack superpotential couplings whose degree is a multiple
of four. It is also interesting that degree~5 couplings were never
allowed.  There is a unique form allowed by string selection
rules, but in the BSL$_A$ class of models this is never a gauge
invariant coupling. It would be of interest to understand this
result more fundamentally; we leave it for future considerations.

\begin{table}[tb]
\caption{\label{tbfrs} \textbf{Restriction of D-flat directions
due to stringent F-flatness}.  The column ``w/o'' indicates the
number of I-monomials that were found without imposing stringent
F-flatness. The column ``w/3'' contains the number that remained
after imposing stringent F-flatness solely with respect to the
degree~3 superpotential couplings. The column ``w/3-9'' provides
the final number of I-monomials that survive our analysis, having
imposed stringent F-flatness up to degree~9.}
\begin{tabular}{|c||c|c|c|} \hline
\parbox{1.2cm}{Pattern} & \parbox{1.2cm}{w/o} & \parbox{1.2cm}{w/3} &
\parbox{1.2cm}{w/3-9} \\ \hline \hline
1.1 & 1486616 & 16283 & 489 \\ 1.2 &
11656 & 188 & 28 \\ \hline 2.1 & 155555 & 1239 & 245 \\ 2.2 &
96932 & 737 & 249 \\ 2.3 & 43884 & 670 & 115
\\2.4 & 5195 & 114 & 12 \\ 2.5 & 12 & 0 & 0
\\ 2.6 & 825 & 9 & 9 \\ \hline 3.1 & 16927 & 80 & 27 \\
3.2 & 2443 & 18 & 10 \\ 3.3 & 9871 & 74 & 22 \\ 3.4 & 1303 & 59 &
41 \\ \hline 4.1 & 17413 & 106 & 26 \\4.2 & 78819 & 513 & 199 \\
4.3 & 14715 & 310 & 163
\\ 4.4 & 26 & 0 & 0 \\ 4.5 & 5126 & 32 & 25 \\
4.6 & 128 & 8 & 5 \\4.7 &   5285 & 15 & 15 \\ 4.8 & 49 & 1 & 1 \\
\hline
\end{tabular}
\end{table}

The second significant result is the extent to which stringent
F-flatness restricts the number of I-monomials. That is to say,
F-flatness is a powerful restriction on flat directions -- perhaps
not a great surprise, but we are able to quantify this in
Table~\ref{tbfrs}. The first column in that Table, in which no
condition of F-flatness is imposed, is the result of Step~1 above,
while the final column is the result of Step~3. It is interesting
that in some models either there is a unique stringently F-flat
direction or no such directions at all, to the order considered
here. A further analysis of these cases is warranted to understand
what is the true nature of the vacuum in these
models;\footnote{The true minimum of the scalar potential may
involve nontrivial cancellations between terms contributing to the
F-terms, so they would correspond to the larger class of flat
directions that are not stringently F-flat.} however, this is
beyond the scope of the present study. It can also be observed
that stringent F-flatness with respect to the degree~3 couplings
is already very limiting. In every model the higher order
couplings only reduce the number of flat directions by a factor of
$\ord{1}$.

But by far the most significant and intriguing result was the
following. We analyzed the first~3 models from each of the~20
BSL$_A$ representation patterns.\footnote{Compare with Table 13
of~\cite{Giedt:2001zw}. We were not able to check all models for
all patterns due to the rather lengthy run-time for the automated
analysis.} For each model that we studied of a given pattern, our
results were identical in terms of the number of couplings of each
degree, the number of initial I-monomials obtained in Step~2, and
the number of I-monomials that survived Step~3.  This provides
further support to what was already indicated in the results
of~\cite{Giedt:2001zw}: the models of a given pattern are in fact
equivalent and that the BSL$_A$ class only contains~20
inequivalent models. This is a drastic reduction from the tens of
thousands that would be expected from naive considerations of all
the different embeddings one can construct that would yield the
same $G_{\rm obs}$. Furthermore, the restrictiveness of F-flatness
is responsible for isolated vacua in $N=1$ models, and is often
invoked in the counting of string vacua.  We wish to emphasize the
relevance of our analysis to ``landscape'' analyses: merely
counting free parameters in some moduli space is not really a
counting of physically distinct vacua.

\mys{Two promising cases} \label{cases}
Two of the twenty patterns were capable of producing a candidate
Majorana neutrino mass as an effective operator of the
form~(\ref{hekr}) along one or more flat directions. Neither
pattern was ultimately able to generate realistic neutrino masses,
however. In this section we will consider each pattern by choosing
a representative model from the set (with the implicit assumption
that all models in a pattern are actually equivalent).

\subsection{Pattern 2.6}
\label{case26}

We will first consider Pattern~2.6 by choosing one of the six
models in the pattern for explicit examination:
Model~2.8.\footnote{The numbering system for the 175~individual
models derives from~\cite{Giedt:2001zw} but is otherwise
irrelevant for our discussion here.} This model is defined by the
following embedding vectors
\begin{eqnarray}
V &=& \frac{1}{3}(-1,-1,0,0,0,2,0,0;2,1,1,0,0,0,0,0) \nonumber \\
a_1 &=& \frac{1}{3}(1,1,-1,-1,-1,-1,0,0;-1,0,0,1,1,1,1,-1)
\nonumber \\
a_3 &=& \frac{1}{3}(0,0,0,0,0,0,2,0;-1,1,1,1,1,1,1,-1)
\label{m28tab1} \end{eqnarray}
and the resulting gauge group is $G = SU(3)\times SU(2) \times
SU(5) \times SU(2) \times U(1)^8$. Our choice for the eight $U(1)$
generators, in terms of the canonical momenta of the $E_8 \times
E_8$ root torus, is given in Table~\ref{m28tab2}. Note that these
generators have been redefined so that only the last generator
$Q_8$ is anomalous, with $\tr Q_8 = 3024$. (This charge is not
canonically normalized; see Table~\ref{m28tab2}.) The spectrum of
chiral superfields and their charges under these eight $U(1)$
factors is given in Table~\ref{m28tab3} of Appendix~\ref{tables}.

\begin{table}[tb]
\caption{\label{m28tab2} \textbf{$U(1)$ charge basis for
Model~2.8.} The eight Abelian factors are defined in terms of the
canonical momenta of the $E_8 \times E_8$ root lattice, with
normalization given in the last column. Canonically normalized
generators $\wh{Q}_a$ are obtained from $\wh{Q} =
Q_a/\sqrt{k_a}$.}
\begin{tabular}{|c||c|c|} \hline
\parbox{0.5cm}{a} & \parbox{4cm}{$Q_a$} & $k_Q/36$ \\ \hline \hline
1 & 6(3,3,-4,-4,-4,0,0,0;0,0,0,0,0,0,0,0) & 132 \\ \hline
2 & 6(2,2,1,1,1,-11,0,0;0,0,0,0,0,0,0,0) & 264 \\ \hline
3 & 6(0,0,0,0,0,0,1,0;0,0,0,0,0,0,0,0) & 2 \\ \hline
4 & 6(0,0,0,0,0,0,0,1;0,0,0,0,0,0,0,0) & 2 \\ \hline
5 & 6(-10,-10,-5,-5,-5,-5,0,0;0,0,0,-12,-12,-12,-12,12) & 2040
\\ \hline
6 & 6(0,0,0,0,0,0,0,0;0,1,1,0,0,0,0,0) & 4 \\ \hline
7 & 6(-10,-10,-5,-5,-5,-5,0,0;-17,0,0,5,5,5,5,-5) & 1428 \\ \hline
8 & 6(-2,-2,-1,-1,-1,-1,0,0;5,0,0,1,1,1,1,-1) & 84
\\ \hline
\end{tabular}
\end{table}


In this model, and for other models in this pattern, we find just
9 I-monomials that survive the requirement of F-flatness to degree
nine in the superpotential (cf.~Table~\ref{tbfrs}). There are~14
effective Majorana masses for candidate right-handed neutrinos
along six of the nine flat directions. These effective mass terms
can be divided into two subsets. In the first we have an effective
Majorana mass at the trilinear order. An example is: \beq {\rm
I-monomial:} && (4,4,6,7,18,35,43,43), \nnn {\rm Eff.~Maj.~mass:}
&& (\underline{4},5,5) \label{emm3} \eeq where we underline the
field(s) that get vev(s) to yield an effective mass coupling;
repeated entries indicate so many powers of the repeated field.
Recall that each field also carries a suppressed family index.
There are six such examples of the coupling $(\underline{4},5,5)$
along six different flat directions.

Using the values for the charges in Table~\ref{m28tab3} it is easy
to show that the combination of fields in the I-monomial
of~(\ref{emm3}) is indeed gauge invariant. Our candidate
right-handed neutrino is thus field \#5, which we will label
$N_5$. But from Table~\ref{m28tab3} we see that the field $N_5$ is
not a complete gauge singlet, but is actually a $(10,2)$
representation of the hidden sector $SU(5)\times SU(2)$ gauge
group. The putative ``Majorana mass'' term is seen to be the
coupling $ \mathbf{5}\; \mathbf{10}\; \mathbf{10}$ of the $SU(5)$
part of this group. What is more, fields charged under both of
these groups are required to obtain vacuum expectation values
along this particular flat direction. This is true of all six flat
directions that allow such a candidate Majorana term. Thus the
right-handed neutrino would need to be identified with a singlet
of the surviving gauge group.

However, the minimal see-saw of~(\ref{simss}) also requires the
coupling of this $N_5$ field to some doublets of the observable
sector $SU(2)$ group. But the presence of $N_5$ in the untwisted
sector of this model prevents any such coupling at the leading
order (as does the requirement of gauge invariance under all the
non-abelian factors). There are no couplings at all in the
superpotential at degree~4 and~5 (see Table~\ref{tbl:bsla}), so
the earliest opportunity for this important Dirac coupling is
degree~6. Of the~642 allowed couplings at degree~6 only three
involve the coupling of the field $N_5$ to doublets of the
observable sector $SU(2)$. These three nonrenormalizable terms
take the form
\begin{eqnarray}
{\rm Coupling (1)}&:& \underline{S_4} N_5 N_5 L_{12} B_{30}
L'_{44}
\\
{\rm Coupling (2)}&:& \underline{S_4} N_5 N_5 L_{40} B_{30}
L'_{22}
\\
{\rm Coupling (3)}&:& \underline{S_4} N_5 N_5 S_{29} B_{30} B_{30}
.
\end{eqnarray}
The fields are labeled according to their type and entry number in
Table~\ref{m28tab3}: $S$ for singlets of all non-abelian groups,
$N$ for the candidate right-handed neutrino, $L$ for doublets of
$SU(2)_{\rm obs}$, $L'$ for doublets of $SU(2)_{\rm hid}$ and $B$
for fields bifundamental under both $SU(2)$ factors. Clearly these
are not standard Dirac mass terms for the field $N_5$.

Even if effective Dirac mass terms that could give rise to the
matrix $(m_D)$ in~(\ref{simss}) were present at degree six in the
superpotential it is still unlikely that we would obtain an
adequate set of neutrino masses. We can estimate the typical scale
of the three light eigenvalues in the following manner. It is
natural to assume that fields such as $S_4$ above obtain a vev
near the scale given by $\xi_{\rm FI}$ in~(\ref{FIterm}). Then the
typical entry in the matrix $m_M$ of~(\ref{simss}) is $r_{\rm FI}
M_{\PL} \sim 0.1 M_{\PL}$, where we have used the information from
Table~\ref{tbl:bsla}. An effective Dirac mass term $m_D$ at degree
six would presumably involve three such vevs, suggesting a set of
light eigenvalues for the matrix~(\ref{simss}) of the form
\begin{equation}
m_{\nu} \sim \frac{(r_{\rm FI}^3 v_u)^2}{r_{\rm FI} M_{\PL}} \sim
r_{\rm FI}^5 \times 10^{-5} \eV
\label{mD28} \end{equation}
where we have used $v_u \sim 100 \GeV$. This suggests neutrino
masses in the nano-eV range -- clearly far too small to fit the
measured squared mass differences.

The second subset of candidate Majorana mass terms involve much
higher-degree superpotential couplings, so one might expect a
better fit to the data. For example, one of the eight remaining
flat direction/Majorana coupling pairs is \beq {\rm I-monomial:}
&& (4,4,7,18,19,27,43,43), \nnn {\rm Eff.~Maj.~mass:} &&
(\underline{7,7,19,27,43,43,43}, 34, 34) . \label{emm9} \eeq Here
the candidate right-handed neutrino is state $N_{34}$ from one of
the twisted sectors of the theory. Yet it is still charged under
the hidden sector gauge group. In this case it is a
$\mathbf{\oline{5}}$ of the hidden $SU(5)$ group. This will again
make it impossible to generate a gauge-invariant Dirac-mass term
at the leading trilinear order. In this case the field $N_{34}$
does not appear {\em in any allowed couplings whatsoever} at
degree~6 in the superpotential, let alone couplings to $SU(2)$
doublets! The next allowed order for such a coupling is then
degree~9, but a dimension-counting argument again gives rise to
the same effective scale as in~(\ref{mD28}) for such a Dirac term
with a degree~nine effective Majorana mass term. We thus conclude
that (i) the required couplings in the minimal see-saw of
Section~\ref{seesaw} do not arise in this model and (ii) the
peculiarities associated with the fact that all candidate
right-handed neutrinos in this model are charged under the hidden
sector $SU(5)$ prevent viable neutrino masses even if they did.

\subsection{Pattern 1.1}
\label{case11}

To exhibit the properties of the candidate neutrino sectors of
Pattern~1.1 we will choose Model~1.2. This model is defined by the
following set of embedding vectors
\begin{eqnarray}
V &=& \frac{1}{3}(-1,-1,0,0,0,2,0,0;2,1,1,0,0,0,0,0) \nonumber \\
a_1 &=& \frac{1}{3}(1,1,-1,-1,2,0,0,0;0,2,0,0,0,0,0,0) \nonumber
\\
a_3 &=& \frac{1}{3}(0,0,0,0,0,0,2,0;-1,0,-1,0,0,0,0,0)
\label{m12tab1} \end{eqnarray}
and the resulting gauge group is $SU(3)\times SU(2) \times SO(10)
\times U(1)^8$. In this model none of the $U(1)$ factors is
anomalous, so we choose a simple basis, in terms of the canonical
momenta of the $E_8 \times E_8$ root torus, for the $U(1)$
generators as given in Table~\ref{m12tab2}. The spectrum of chiral
superfields and their charges under these eight $U(1)$ factors is
given in Table~\ref{m12tab3} of Appendix~\ref{tables}.

The lack of an anomalous $U(1)_X$ suggests that we can no longer
assume vevs for fields in a particular flat direction are at the
scale $\xi_{\rm FI}$. Nevertheless, the existence of flat
directions (or nearly flat directions) allows us to consistently
choose scalar fields to have large
vevs~\cite{Cvetic:1997ky,Cleaver:1997nj}. The determination of the
exact size of these vevs requires minimization of the effective
scalar potential for these D-moduli.

\begin{table}[tb]
\caption{\label{m12tab2} \textbf{$U(1)$ charge basis for
Model~1.2.} The eight Abelian factors are defined in terms of the
canonical momenta of the $E_8 \times E_8$ root lattice, with
normalization given in the last column.}
\begin{tabular}{|c||c|c|} \hline
\parbox{0.5cm}{a} & \parbox{6cm}{$Q_a$} & \parbox{1.2cm}{$k_Q/36$} \\ \hline \hline
1 & 6(-3,-3,2,2,2,0,0,0;0,0,0,0,0,0,0,0) & 60 \\ \hline
2 & 6(1,1,1,1,1,0,0,0;0,0,0,0,0,0,0,0) & 10 \\ \hline
3 & 6(0,0,0,0,0,1,0,0;0,0,0,0,0,0,0,0) & 2 \\ \hline
4 & 6(0,0,0,0,0,0,1,0;0,0,0,0,0,0,0,0) & 2 \\ \hline
5 & 6(0,0,0,0,0,0,0,1;0,0,0,0,0,0,0,0) & 2 \\ \hline
6 & 6(0,0,0,0,0,0,0,0;0,0,1,0,0,0,0,0) & 2 \\ \hline
7 & 6(0,0,0,0,0,0,0,0;0,1,0,0,0,0,0,0) & 2 \\ \hline
8 & 6(0,0,0,0,0,0,0,0;1,0,0,0,0,0,0,0) & 2 \\ \hline
\end{tabular}
\end{table}


As shown in Table~\ref{tbfrs}, a total of 489~I-monomials that
satisfy stringent F-flatness to degree~9 were found.  From these,
and the 3,427,239 couplings that appear in Table~\ref{spcpn} for
this model, 42 instances of effective Majorana mass couplings were
found along 18~of the~489 flat directions.  For brevity, we do not
enumerate all of these flat directions and effective Majorana mass
couplings, but confine ourselves to a discussion of some
representative examples. Remarkably, though there are nominally
42~different pairs of flat direction/effective Majorana operator,
these pairs form patterns that repeat themselves -- the labels on
the fields may change, but the representations and structure do
not. Thus a very small number of actual possibilities exist. All
of the Majorana couplings/candidate neutrinos fall into one or the
other of the two cases given below.


For our first example, let us consider the flat direction
characterized by the following pair of invariant and Majorana
operator:
\beq {\rm I-monomial:} && (2,2,3,3,8,8,34,46,61,77), \nnn {\rm
Eff.~Maj.~mass:} && (\underline{2,3,8,8,34,46},74,74) .
\label{emm1} \eeq
It can be seen from the spectrum of Table~\ref{m12tab3} that all
the fields getting vevs in the flat direction are non-abelian
singlets. The candidate right-handed neutrino is field \#74 which
we denote $N_{74}$, using the notation described above. This flat
direction leaves two $U(1)$ factors unbroken: that is, there are
two linear combinations of $U(1)$ generators such that the charges
of all the fields in the first line of~(\ref{emm1}) can be made
simultaneously zero. Therefore, if the Standard Model hypercharge
generator can be identified with one of these linear combinations
the gauge invariance requirement of the superpotential terms will
automatically enforce $Q_Y^{N_{74}} = 0$.

This candidate right-handed neutrino does not appear in any
allowed trilinear superpotential coupling.  A careful analysis of
the degree~4 superpotential couplings shows that the flat
direction in \myref{emm1} does not produce an effective Dirac
coupling. This conclusion is also true with respect to the 21,329
couplings at degree~6. Since the $U(1)$'s are non-anomalous in
this case, we cannot estimate the scale of the vevs in the flat
direction, other than that it is below~$M_{\PL}$. If their typical
scale is $\lang S \rang/M_{\PL} \equiv c < 1$, then the light
neutrino eigenvalues will be of the order
\begin{equation}
m_{\nu} \sim \frac{(c^{d-3} v_u)^2}{c^6 M_{\PL}} \sim c^{2d-12}
\times 10^{-5}\eV ,
\label{xxx} \end{equation}
where $d$ is the degree of the effective Dirac mass term. Thus an
acceptable mass would require $d$ smaller than~6.

A much more promising case is the one characterized by the
following invariant:
\beq {\rm I-monomial:} && (3,3,8,21,22,29,46,72) .  \label{emm2}
\eeq
Once again, it can be seen from the spectrum of
Table~\ref{m12tab3} that all the fields getting vevs in the flat
direction are non-abelian singlets. Along this direction there are
two effective Majorana mass operators, one at degree~6 and the
other at degree~8
\beq {\rm Eff.~Maj.~mass (1):} && (\underline{8,22,46,72},9,9)
\nnn {\rm Eff.~Maj.~mass (2):} &&
(\underline{3,3,8,22,46,72},13,13) . \label{Maj} \eeq
The two Majorana operators differ only by the insertion of two
untwisted sector fields $S_3$. The candidate right-handed
neutrinos are thus $N_9$ and/or $N_{13}$. Along this flat
direction three $U(1)$ factors remain unbroken to low energies,
and all linear combinations of these three $U(1)$'s allow
$Q_Y^{N_{9,13}} = 0$.

Once again, identifying either field as a bona fide neutrino
requires looking for the requisite Dirac couplings to $SU(2)$
doublets. Here the outlook is much brighter: there are several
couplings involving $SU(2)$ doublets and both $N_9$ and $N_{13}$
at degree~3 and degree~4 in the superpotential. In fact, each
admits two such couplings
\begin{equation}
(A) \;\; \lbr \begin{array}{c} N_9 L_{36} L_{64} \\
\underline{S_3} N_{13} L_{36} L_{64} \end{array} \right. \quad (B)
\;\; \lbr \begin{array}{c} N_9 L_{52} L_{71} \\ \underline{S_3}
N_{13} L_{52} L_{71} \end{array} \right. ,
\label{Dirac} \end{equation}
where we use $L$ to denote doublets of $SU(2)$. At this stage we
are not yet in a position to distinguish lepton doublets from
up-type Higgs doublets as we have not yet identified other Yukawa
interactions or designated a unique hypercharge assignment. Thus
we use a common notation for both doublets.

We thus appear to have the two essential ingredients for forming
the matrix of couplings in~(\ref{simss}) and, in fact, we have the
potential for embedding the entire leptonic sector of the MSSM
superpotential. For instance, if we make the identification
$L_{36} = L$ and $L_{64} = H_u$, then we find trilinear couplings
of the form
\begin{equation}
W \ni \lambda_1 L_{36} L_{5} S_{60} + \lambda_2 L_{64} L_{5}
S_{38} .
\label{alllep} \end{equation}
If we further identify $L_{5} = H_d$ we see that $S_{60}$ could
play the role of the right-handed charged lepton field $E^c$ while
$S_{38}$ could generate an effective $\mu$-term through some
additional low-energy dynamics. If we were instead to make the
identification $L_{36} = H_u$ and $L_{64} = L$ we would merely
need to exchange the interpretation of the fields $S_{38}$ and
$S_{60}$. There are several possible systems such
as~(\ref{alllep}) for either choice of couplings~(A) or~(B)
of~(\ref{Dirac}).

Of course this discussion assumes that the correct hypercharge can
simultaneously be assigned to each of these fields along a
particular surviving $U(1)$ combination. In many cases this is
indeed possible. For example, in the system given
by~(\ref{alllep}) and the identification $\lbr N_9, L_{36},
L_{64}, L_5, S_{60}, S_{38} \rbr \leftrightarrow \lbr N, L, H_u,
H_d, E^c, S \rbr$ the particular linear combination of $U(1)$
factors that gives rise to the correct hypercharge assignments
$\lbr 0 , -1/2, 1/2, -1/2, 1, 0 \rbr$ is given by
\begin{eqnarray}
U(1)_Y& =& -\frac{7}{180}U(1)_1 -\frac{1}{30}U(1)_2 +
\frac{1}{6}U(1)_4  \nonumber \\
 & & +\frac{1}{4}U(1)_6 -\frac{1}{4}U(1)_7 + \frac{1}{12}U(1)_8 .
\label{hyper} \end{eqnarray}
This also accommodates the quarks of the MSSM. However, the
hypercharge normalization is $k_Y=91/6$ rather than the
$SU(5)$-based GUT value of $k_Y=5/3$. This is not consistent
(perturbatively) with the observed couplings, even allowing for
the effects of additional matter states in the running of the
gauge couplings. However, our purpose in this study is to focus on
the neutrino sector and examine how many, if any, of flat
directions allow a minimal seesaw, irrespective of whether they
are fully realistic in other ways.

So far, so good. We cannot say anything very definite about the
typical scale of the vevs $\lang S_i \rang$ that give rise to the
effective Majorana couplings in~(\ref{Maj}) since there is no
anomalous $U(1)$ factor in the model. But given that Dirac mass
terms can arise at degree three or four, a scale somewhere between
the GUT and string scales for these vevs would be welcome. This is
not impossible to imagine, since the Standard Model singlet fields
involved in the flat direction~(\ref{emm2}) do couple to several
doublets of $SU(2)$ and triplets/anti-triplets of $SU(3)$ at the
trilinear order. These are just the sorts of ingredients that can
give rise to a high, radiatively-generated intermediate
scale~\cite{Cleaver:1997nj}. It would be tempting, then, to
declare victory and begin to calculate the possible mass textures
for both the Majorana and Dirac matrices -- perhaps by assuming
only third generation Higgs fields and singlets in~(\ref{emm2})
receive vevs so that selection rules would then enforce texture
zeros in the effective mass matrices.

But this would be premature. To begin with, we should note that
there are no quark masses in this model at the leading order; by
placing the up-type Higgs doublet in the twisted sector (as all of
the examples in this class require to generate the neutrino Dirac
mass) it becomes impossible to couple it to the (untwisted sector)
quark doublet at the trilinear order.\footnote{We do not consider
the possibility of different families of up-type Higgs doublets
involved in generating the neutrino and quark masses,
respectively.} The desired quark masses do not appear at degree~4
either, and at degree six we find only one new coupling involving
quark doublets
\begin{equation}
W^6 \ni \underline{S_3 S_8 S_{72}} Q_1 Q_1 D_{56}
\label{QQD} \end{equation}
where three of the fields are participants in the flat direction,
the quark doublet is repeated, and the coupling is to field \#56
which is a $\mathbf{3}$ of $SU(3)$. This is certainly not the
quark sector of the Standard Model, and clearly there are no GUT
relations between the neutrino Yukawa interactions and those of
the up-type quarks. But our analysis was based on answering the
sole question of whether the minimal see-saw can be found in an
explicit string construction, so we will not consider the quarks
further.

Of greater concern is the redundancy evidenced by the multiple
neutrino candidates and multiple Higgs candidates in this example.
Can these extra states be projected out of the light spectrum
along the flat direction, perhaps leaving only one of the sets of
couplings in~(\ref{Dirac})? Do the remaining light states, and in
particular the candidate right-handed neutrinos, mix with one
another? To fully understand the nature of neutrino masses in this
set of examples a thorough analysis that considers all the
relevant fields of the system must be performed. When we do so we
will see that our earlier enthusiasm for this set of flat
directions and couplings was misplaced.

A careful consideration of all degree~3 and~4 couplings in the
superpotential indicates that many of the extra $SU(2)$ doublets
(and, incidentally, all of the exotic $\mathbf{3}$'s of $SU(3)$)
are projected out of the spectrum -- a welcome development. For
example we find the couplings
\begin{eqnarray}
W &=& \lambda_1 \underline{S_{21}} L_{49} L_{70} + \lambda_2
\underline{S_{22}} L_{12} L_{24} \nonumber \\ & & + \lambda_3
\underline{S_{29}} L_{51} L_{80} + \lambda_4 \underline{S_{46}}
L_{47} L_{48}
\label{projL} \end{eqnarray}
which eliminates all the possible combinations of $W = \lambda N L
H_u$ associations but the two listed in~(\ref{Dirac}). As there is
no reason to choose $N_9$ versus $N_{13}$ as our right-handed
neutrino, we must therefore conclude that the neutrino sector of
this theory involves at least two {\em species} of neutrino, each
with three generations. So too we must accept two species of
lepton doublets, and without loss of generality we may choose them
to be $L_{36}$ and $L_{52}$, with fields \#64 and \#71 being two
species of up-type Higgs doublets.

So this model does not give rise to a minimal see-saw after all.
In fact, there are terms that mix our fields with Dirac couplings
($N_9$ and $N_{13}$) with other Standard Model singlets that do
not. We will refer to these additional states with the notation
$\wtd{N}$. In particular we have the couplings
\begin{eqnarray}
W_{\rm mix} & = & \lambda \underline{S_8}N_9 \wtd{N}_{14} +
\lambda \underline{S_{22}}N_9 \wtd{N}_{27} \nonumber \\ & & +
\lambda \underline{S_{72}}N_9 \wtd{N}_{50} + \lambda
\underline{S_{46}}N_9 \wtd{N}_{81} \nonumber \\
 & & + \lambda
\underline{S_3 S_{8}}N_{13} \wtd{N}_{14} + \lambda \underline{S_3
S_{22}}N_{13} \wtd{N}_{27} \nonumber \\ & & + \lambda
\underline{S_3 S_{72}}N_{13} \wtd{N}_{50} + \lambda \underline{S_3
S_{46}}N_{13} \wtd{N}_{81}
\label{mix} \end{eqnarray}
which generate an extended see-saw structure. As mentioned in the
introduction, this is not an uncommon feature of explicit string
constructions, in part because of the large numbers of Standard
Model singlets that are generally present. Nor need it imply that
small neutrino masses are impossible to obtain.

In this particular example the effective neutrino system mass
matrix is given in block matrix form by
\begin{equation}
\begin{array}{c} ( \nu_L \;\; \wtd{N} \;\; N ) \\ \\
\end{array}
\( \begin{array}{ccc}  0 & 0 & A
\\ 0 & 0 & B \\ A & B & C
\end{array} \)
\( \begin{array}{c}\nu_L \\ \wtd{N} \\ N  \end{array} \) ,
\label{massblock} \end{equation}
defined with the basis sets
\begin{eqnarray}
\nu_L &=& \lbr (\nu_L)_{36} , \; (\nu_L)_{52}\rbr, \nonumber \\
\wtd{N}& =& \lbr \wtd{N}_{14} , \; \wtd{N}_{27}, \; \wtd{N}_{50},
\; \wtd{N}_{81} \rbr, \nonumber \\
N &=& \lbr N_9, \; N_{13} \rbr .
\label{sets} \end{eqnarray}
The individual submatrices in~(\ref{massblock}) are
\begin{eqnarray}
A &=& \(\begin{array}{cc} \lang (H_u)_{64} \rang & \lang S_3
(H_u)_{64} \rang \\ \lang (H_u)_{71} \rang & \lang S_3 (H_u)_{71}
\rang \end{array} \) \nonumber \\
B &=& \( \begin{array}{cc} \lang S_8 \rang & \lang S_3 S_8 \rang
\\ \lang S_{22} \rang & \lang S_3 S_{22} \rang \\ \lang S_{72} \rang &
\lang S_3 S_{72} \rang \\ \lang S_{46} \rang & \lang S_3 S_{46}
\rang
\end{array} \) \nonumber \\
C &=& \( \begin{array}{cc} \lang S_8 S_{22} S_{46} S_{72} \rang &
0 \\ 0 & \lang S_3^2 S_8 S_{22} S_{46} S_{72}\rang \end{array} \)
\label{secmat} \end{eqnarray}
with the general expectation that the pattern of vevs would be
such that $A \ll C \ll B$. But the matrix~(\ref{massblock}) has
vanishing determinant and gives rise to three precisely massless
eigenvalues; the ``see-saw'' serves only to split the masses of
the very heavy eigenstates. This mechanism, in which the addition
of off-diagonal terms in an extended right-handed sector destroys
what appeared to be a successful construction, could easily occur
in other constructions and should be checked for.

We might hope to populate some of the zero blocks
in~(\ref{massblock}) to salvage this example (though we are
already far from a minimal see-saw), but there are no Dirac
couplings of doublets $L_{36}$ or $L_{64}$ to the fields $\wtd{N}$
at degree~3,~4 or~6.  There are also no couplings at degree $d<6$
that couple the fields in the $\wtd{N}$ system to themselves --
that is, the entire matrix of values represented by the $(2,2)$
entry in~(\ref{massblock}) is vanishing to this degree. We could
imagine expanding the system yet again, and looking for couplings
of an expanded $\wtd{N}$ system where effective mass terms can
arise, say when one of the remaining $U(1)'$ factors is
spontaneously broken. But here again by considering all allowed
operators of this form at degree $d<6$ in the superpotential, we
find the determinant of this submatrix always vanishes, indicating
vanishing eigenvalues for the full matrix~(\ref{massblock}). By
arguments similar to those that gave rise to~(\ref{mD28})
and~(\ref{xxx}) it is easy to see that realistic masses for the
light neutrinos would require either a Dirac-type coupling of the
$\wtd{N}$ to the $\nu_L$ fields or Majorana masses for the
$\wtd{N}$ fields at no higher degree than the trilinear order.
Thus we have succeeded in finding the right operators for the
$(\nu_L,\;N)$ system in isolation, but by considering the full
lepton system we find that we have failed to account for the
finite and small neutrino masses observed in nature.

\subsection{Why so many zeros?}
It is of interest to understand why the zeros have appeared in the
mass matrices~(\ref{massblock}) and~(\ref{secmat}). Are they
exact? Are they the consequence of a symmetry? If they are not
exact, at what order do nonzero contributions first appear, and
what would be the effects?  We have studied these questions, and
here we will summarize the answers.

Since the fields of type $N$ and $\wtd{N}$ are $U(1)_Y$ neutral,
the zeros do not follow from this symmetry.  However, one might
ask: Are the zeros explained by the two extra $U(1)$'s that
survive along the flat direction~(\ref{emm2})? A useful basis for
this $U(1)^2$ subgroup consists of the following (canonically
normalized) generators:
\begin{eqnarray}
A &=& -\frac{4}{15} Q_1 + \frac{67}{60} Q_2 - \frac{15}{4} Q_3 +
\frac{23}{12} Q_4 \nonumber \\
 & & - \frac{15}{4} Q_5 - \frac{11}{6}
Q_6 + \frac{11}{6} Q_7 + \frac{17}{3} Q_8 \\
B &=& -\frac{119}{30} Q_1 - \frac{139}{60} Q_2 + \frac{91}{12} Q_3
- \frac{43}{12} Q_4 \nonumber \\
 & & + \frac{91}{12} Q_5 - \frac{55}{6} Q_6 + \frac{55}{6} Q_7 + 2 Q_8 .
\end{eqnarray}
The $N$ and $\wtd{N}$ fields are all neutral with respect to these
generators. It follows that none of the zeros in the mass matrix
are a consequence of gauge invariance.

Further investigation finds that there are discrete symmetries
that survive along the flat direction that we study here.  Along
the flat direction~(\ref{emm2}) there is a breaking $U(1)^8 \to
U(1)^3$. However, a discrete subgroup of the broken $U(1)^5$
survives, when it is combined with the the trialities of the
original theory. This subgroup is found by demanding that the
fields in~(\ref{emm2}) are left invariant. However, we find that
these surviving discrete symmetries do not explain the zeros
either.

In fact, we find that the zeros are not exact, but are violated by
high order terms.  For instance, mass terms of the form $m_{AB}
\wtd{N}_A \wtd{N}_B$ are allowed by all symmetries and appear with
$m_{AB}$ being a degree~10 polynomial of the fields~(\ref{emm2}).
This translates into Majorana mass terms of order $10^8$ GeV or
less. By arguments similar to those made above, such entries would
only generate neutrino masses of order $m_\nu \sim c^{8} \times
10^{-5}$ eV, where $c=S/M_{P} < 1$.  Thus $m_\nu < 10^{-13}$ eV
for $c < 0.1$, a negligible effect.

In summary, the zeros that we find are not exact, but they may as
well be, since the allowed violations of them are of such high
order. This is a consequence of the fact that we have studied all
terms of the superpotential to a very high order.

\mys{Conclusions} \label{conc}

Our systematic search of the BSL$_A$ class of otherwise
phenomenologically promising, top-down constructions of the
heterotic string failed to reveal a minimal see-saw mechanism.
This despite our placing the existence of an effective Majorana
mass operator at the fore of our search -- a search through
literally millions of superpotential couplings along thousands of
flat directions in a class with dozens of potentially realistic
models. What are we to conclude from this null result?

It may be that our search, as computationally intensive as it was,
was not sufficiently broad to find the elusive couplings. One
might imagine generalizations in which several flat-directions, or
I-monomials, are simultaneously ``turned on.'' This is, of course,
a very real physical possibility that might allow for more
effective mass terms (since more vevs are non-vanishing). But it
will also tend to be even more severely constrained by F-flatness
conditions. We might, in addition explore directions that are not
stringently F-flat; that is, directions where various nonvanishing
terms cancel to give $\vev{\p W/\p \phi_i}=\vev{W}=0$, or in which
the flatness breaking terms are small enough to be harmless (as
would occur when the vevs are at an intermediate scale). This
would allow for more of the D-flat directions to survive in the
generalization of Step~3 of Section~\ref{flat} above, yielding
more effective mass couplings. This would be an interesting
starting point for a future study, but it should be clear from the
extensive discussion in Sections~\ref{sca} and~\ref{cases} that
the computational demands would grow significantly if one were to
depart from the simple rules we followed.

But one might have thought that if the minimal see-saw -- or
indeed a see-saw mechanism at all -- is the answer to the problem
of small neutrino masses then it should arise with great
frequency, even in a simple search such as ours. That it did not
might be merely a reflection of the peculiarities of the $Z_3$
orbifold itself, or perhaps of orbifold constructions more
generally. Alas, we are unable to address such a question since
the starting point to this work does not even exist for other
orbifolds, much less more general heterotic constructions. Yet the
$Z_3$ has been well-studied in the past precisely because it has
so many other desirable features. That it does not seem to possess
a minimal see-saw mechanism should perhaps give us pause.

Thus we might consider again the most successful case we found in
our study, that of Pattern~1.1 as described in
Section~\ref{case11}. Here we find a structure that is not
minimal, but of the extended see-saw variety. While this
particular example was unable to give mass to the neutrino
eigenstates, it may well be that string theory prefers, or at
least can accommodate, neutrino mass mechanisms that are not
minimal. We dedicated our analysis to the search for couplings of
the form~(\ref{hekr}), but we might instead have considered the
more general extended form
\beq \vev{S_1 \cdots S_{n-2}} N N' \label{heks} \eeq
where $N$ and $N'$ are different {\it species} in the sense
defined in~\cite{Giedt:2001zw}.  That is, $N$ and $N'$ differ by
something more than the third fixed point label that corresponds
to the 3-fold degeneracy of this class of models. Needless to say,
a systematic search would immediately have to confront the
enormous increase in combinatorics involved in such couplings.

Generically, \myref{heks} corresponds to an effective theory with
6~right-handed neutrinos. Given our difficulty in finding
couplings of the form~\ref{hekr} we might expect such an extended
mass matrix to have vanishing diagonal entries. Then achieving
appropriate neutrino masses would require a mass matrix of the
form
\beq W_{\stxt{eff}} = ( \nu_i, N_i, N'_i)
\begin{pmatrix} 0 & (m_D)_{ij} & (m'_D)_{ij} \cr (m_D)_{ji} & 0 &
(m_M)_{ij} \cr (m'_D)_{ji} & (m_M)_{ji} & 0 \end{pmatrix}
\begin{pmatrix} \nu_j \cr N_j \cr N'_j \cr \end{pmatrix}
\label{extss} \eeq
It is also worth noting that the $m_D, m'_D$ can have a certain
amount of texture zeros and still give all neutrinos mass.
However, nonzero elements for $m_D$ and $m'_D$ would both have to
be present. The see-saw still gives 3 light flavors of mass
$\ord{m_D^2/m_M}$ and now 6 heavy flavors of mass $\ord{m_M}$. An
extended see-saw model, though of a somewhat different form
from~(\ref{extss}), has previously appeared in free-fermionic
constructions~\cite{Faraggi:1990it}, as noted in the Introduction.

It may also be that the standard see-saw ideas are not the answer
to small neutrino masses. Twelve of the twenty cases listed in
Table~\ref{tbl:bsla} contain fields which are bifundamental under
two different $SU(2)$ factors. If these groups were to be broken
to the diagonal subgroup, such states would be effective triplets
under the surviving $SU(2)$. If they were sufficiently massive,
they may form the basis for a Type~II see-saw
mechanism~\cite{typeii}. An investigation of this possibility in
the $Z_3$ orbifold is underway~\cite{triplet}. Dirac-type
couplings of the form $N L H_u$ are much more common than
effective Majorana mass-type couplings. It may therefore simply be
that neutrinos are Dirac particles, with neutral lepton Yukawa
couplings only arising at higher order in the effective
superpotential. These terms would have to be of extremely high
order if the relevant vevs are close to $M_{\PL}$, or could be as
low as degree~4 if the vevs are at some intermediate
scale~\cite{Cleaver:1997nj,Langacker:1998ut}. This is contrary to
most theoretical prejudice, but it is certainly a logical
possibility. One of the lessons of this study is that in explicit
string constructions couplings, such as those leading to Majorana
masses, are determined to a greater degree by the underlying
theory and less by esoteric reasoning (naturalness, elegance,
simplicity, etc.). While we would have preferred to have found
models with flat directions where the simple see-saw works, we
therefore regard our null result as significant.

Our principal result also demonstrates the power of the underlying
string theory.  It is remarkable that something as simple as
\myref{hekr} is not possible along a flat direction. Certainly one
would not expect this from a ``bottom-up'' perspective.  This
result is a consequence of the wealth of symmetry constraints that
arise from the underlying theory. These are features that one is
unlikely to have ever guessed. We feel that this demonstrates the
importance of attempting to connect effective field theory
model-building with an underlying theory -- or in modern parlance,
an ``ultraviolet completion.''

We have found further evidence that there are only 20 inequivalent
models in the BSL$_A$ class.  This drastic reduction from a naive
estimate -- based on the number of seemingly different embedding
vectors -- can be given the following interpretation. It shows
that surveys of classes of string constructions can be done; and,
that they can produce meaningful results much the way that a scan
over some significant section of parameter space in an effective
field theory (such as the MSSM) can have meaning. We find this
encouraging, because it hints that qualitative impressions gained
in such a survey are a good guide to effective field theory
model-building.

We last note the importance of having a useful query in mind when
surveying explicit string constructions. The unique nature of the
coupling~(\ref{hekr}) made it an extremely powerful tool in
directing our attention to only a handful of promising cases from
a vacuum space that (prior to the pioneering work of a number of
theorists) looked to include hundreds of thousands of possible
vacuum configurations, with thousands of flat directions to study
in each one.

\begin{acknowledgments}
J.G.~was supported by the National Science and Engineering
Research Council of Canada and the Ontario Premier's Research
Excellence Award. G.K.~thanks the U.S.~Department of Energy for
support. P.L.~and B.N.~were supported by the U.S.~Department of
Energy under Grant No.~DOE-EY-76-02-3071. G.K.~and~B.N. thank the
Aspen Center for Physics for hospitality during certain portions
of this work.
\\
\end{acknowledgments}

\myappendix

\mys{Spectra for promising cases} \label{tables}
\begin{table}[t]
\caption{\label{m28tab3} \textbf{Partial spectrum of chiral matter
for Model~2.8}. Chiral superfields are grouped by sector of the
string Hilbert space. Irreducible representations under the
non-Abelian gauge group $SU(3)\times SU(2) \times SU(5)_H \times
SU(2)_H$ is given, along with charges under the eight $U(1)$
factors. Note that $Q_8$ is the anomalous $U(1)_X$. }
\begin{tabular}{|c|l|r|r|r|r|r|r|r|r|}
\hline  & Irrep.
 & \parbox{0.6cm}{$Q_1$} & \parbox{0.6cm}{$Q_2$} & \parbox{0.6cm}{$Q_3$}
 & \parbox{0.6cm}{$Q_4$} & \parbox{0.6cm}{$Q_5$} & \parbox{0.6cm}{$Q_6$}
 & \parbox{0.6cm}{$Q_7$} & \parbox{0.6cm}{$Q_8$} \\ \hline \hline
\multicolumn{10}{|c|}{sector:  untwisted} \\
\hline 1 & $(1, 2, 1, 1)_{0}$ & $18$ & $-54$ & $0$ & $0$ & $-90$ &
$0$ & $-90$ & $-18$ \\
2 & $(3, 2, 1, 1)_{0}$ & $6$ & $-18$ & $0$ & $0$ & $90$ & $0$ &
$90$ & $18$ \\
3 & $(\bar 3, 1, 1, 1)_{0}$ & $-24$ & $72$ & $0$ & $0$ & $0$ & $0$
& $0$ & $0$ \\
4 & $(1, 1, 5, 1)_{0}$ & $0$ & $0$ & $0$ & $0$ & $72$ & $0$ & $72$
& $-36$ \\
5 & $(1, 1, 10, 2)_{0}$ & $0$ & $0$ & $0$ & $0$ & $-36$ & $0$ &
$-36$ & $18$ \\ \hline
\multicolumn{10}{|c|}{sector: (-1,-1)} \\ \hline
6 & $(1, 1, 1, 1)_{0}$ & $-12$ & $14$ & $2$ & $0$ & $-190$ & $0$ &
$-54$ & $6$ \\
7 & $(1, 1, 1, 2)_{0}$ & $-12$ & $14$ & $2$ & $0$ & $-10$ & $0$ &
$-78$ & $-24$ \\
%
\hline \multicolumn{10}{|c|}{sector: (-1,0)} \\ \hline
12 & $(1, 2, 1, 1)_{0}$ & $-30$ & $2$ & $0$ & $0$ & $-70$ & $-2$ &
$-36$ & $18$ \\
%
%
\hline \multicolumn{10}{|c|}{sector:  (-1,1)} \\ \hline
18 & $(1, 1, 1, 2)_{0}$ & $-12$ & $14$ & $-2$ & $0$ & $-70$ & $2$
& $-36$ & $-24$ \\
19 & $(1, 1, 1, 1)_{0}$ & $-12$ & $14$ & $-2$ & $0$ & $-70$ & $-4$
& $-138$ & $6$ \\
%
%
%
22 & $(1, 1, 1, 2)_{0}$ & $6$ & $-40$ & $1$ & $-3$ & $20$ & $2$ &
$54$ & $-6$ \\
%
%
\hline \multicolumn{10}{|c|}{sector: (0,-1)} \\ \hline
%
27 & $(1, 1, 1, 1)_{0}$ & $-12$ & $14$ & $2$ & $0$ & $-10$ & $6$ &
$24$ & $30$ \\
\hline \multicolumn{10}{|c|}{sector: (0,0)} \\ \hline
%
%
29 & $(1, 1, 1, 1)_{0}$ & $-12$ & $-52$ & $0$ & $0$ & $20$ & $4$ &
$-48$ & $24$ \\
30 & $(1, 2, 1, 2)_{0}$ & $6$ & $26$ & $0$ & $0$ & $-10$ & $-2$ &
$24$ & $-12$ \\
31 & $(1, 2, 1, 1)_{0}$ & $6$ & $26$ & $0$ & $0$ & $-10$ & $4$ &
$-78$ & $18$ \\
\hline \multicolumn{10}{|c|}{sector: (0,1)} \\ \hline
%
%
35 & $(1, 1, 1, 1)_{0}$ & $-12$ & $14$ & $-2$ & $0$ & $110$ & $2$
& $-60$ & $30$ \\ \hline \multicolumn{10}{|c|}{sector:  (1,-1)}
\\ \hline
%
38 & $(\bar 3, 1, 1, 1)_{0}$ & $0$ & $-22$ & $2$ & $0$ & $-10$ &
$0$ & $-78$ & $18$ \\
40 & $(1, 2, 1, 1)_{0}$ & $-12$ & $14$ & $-1$ & $3$ & $-10$ & $0$
& $-78$ & $18$ \\
%
\hline \multicolumn{10}{|c|}{sector: (1,0)} \\ \hline
%
%
43 & $(1, 1, \bar 5, 1)_{0}$ & $24$ & $-28$ & $0$ & $0$ & $8$ &
$-2$ & $42$ & $0$ \\
44 & $(1, 1, 1, 2)_{0}$ & $24$ & $-28$ & $0$ & $0$ & $80$ & $4$ &
$12$ & $-6$ \\
%
\hline \multicolumn{10}{|c|}{sector: (1,1)} \\ \hline
50 & $(1, 2, 1, 1)_{0}$ & $-12$ & $14$ & $1$ & $-3$ & $-70$ & $2$
& $-36$ & $18$ \\
\hline
\end{tabular}
\end{table}

In this Appendix we provide a partial spectrum for Models~2.8
and~1.2. These are the two examples from Pattern~2.6 and~1.1,
respectively, that were chosen for detailed study of the neutrino
sector in the text. Each species of chiral superfield carries a
sequential numerical label. For each species there is a three-fold
replication of generations. Fields are identified by their
irreducible representation under the non-abelian parts of $SU(3)
\times SU(2) \times G_{\rm hid}$ and by their charges under the
Abelian gauge factors. We group states by sector of the string
Hilbert space, beginning with the untwisted sector and followed by
each of the twisted sectors. These twisted states are labeled by
two integers, indicating the fixed point location in each of the
first two compact complex planes (the location in the third plane
being the three-fold degeneracy that gives rise to the three
generations). These integers may take the value 1, 0 or -1 in our
convention. The subscript following the irreducible representation
label denotes the string oscillator number, if any, for the state.

To keep these tables manageable we have included only those states
which are doublets of the Standard Model $SU(2)$ factor or are otherwise
mentioned in the text. Complete tables of spectra can be obtained
from the authors by request.


\begin{table}[bt]
\caption{\label{m12tab3} \textbf{Partial spectrum of chiral matter
for Model~1.2}. Notation is identical to that of
Table~\ref{m28tab3}. This model has no anomalous $U(1)$ factor. }
\begin{tabular}{|c|l|r|r|r|r|r|r|r|r|}
\hline  & Irrep.
 & \parbox{0.6cm}{$Q_1$} & \parbox{0.6cm}{$Q_2$} & \parbox{0.6cm}{$Q_3$}
 & \parbox{0.6cm}{$Q_4$} & \parbox{0.6cm}{$Q_5$} & \parbox{0.6cm}{$Q_6$}
 & \parbox{0.6cm}{$Q_7$} & \parbox{0.6cm}{$Q_8$} \\ \hline \hline
\multicolumn{10}{|c|}{sector:  untwisted} \\ \hline
1 & $(3, 2, 1)_{0}$ & $6$ & $-12$ & $0$ & $0$ & $0$ & $0$ & $0$ &
$0$ \\
2 & $(1, 1, 1)_{0}$ & $0$ & $0$ & $-6$ & $0$ & $-6$ & $0$ & $0$ &
$0$ \\
3 & $(1, 1, 1)_{0}$ & $0$ & $0$ & $-6$ & $0$ & $6$ & $0$ & $0$ &
$0$ \\
4 & $(1, 1, 1)_{0}$ & $0$ & $0$ & $0$ & $0$ & $0$ & $-6$ & $0$ &
$6$ \\
\hline \multicolumn{10}{|c|}{sector: (-1,-1)} \\ \hline
5 & $(1, 2, 1)_{0}$ & $18$ & $4$ & $-2$ & $2$ & $0$ & $-2$ & $-2$
& $0$ \\
%
%
%
8 & $(1, 1, 1)_{0}$ & $0$ & $10$ & $4$ & $2$ & $0$ & $-2$ & $-2$ &
$0$ \\
9 & $(1, 1, 1)_{0}$ & $0$ & $-5$ & $-5$ & $-1$ & $3$ & $-2$ & $-2$
& $0$ \\
%
%
%
12 & $(1, 2, 1)_{0}$ & $-18$ & $1$ & $1$ & $-1$ & $3$ & $-2$ &
$-2$ & $0$ \\
13 & $(1, 1, 1)_{1}$ & $0$ & $-5$ & $1$ & $-1$ & $-3$ & $-2$ &
$-2$ & $0$ \\
14 & $(1, 1, 1)_{0}$ & $0$ & $-5$ & $1$ & $-1$ & $-3$ & $4$ & $4$
& $0$ \\
\hline \multicolumn{10}{|c|}{sector: (-1,0)} \\ \hline
%
%
%
%
%
%
%
21 & $(1, 1, 1)_{0}$ & $0$ & $-5$ & $1$ & $3$ & $-3$ & $-4$ & $-2$
& $-2$ \\
22 & $(1, 1, 1)_{0}$ & $0$ & $-5$ & $1$ & $3$ & $-3$ & $2$ & $4$ &
$-2$ \\
%
%
\hline \multicolumn{10}{|c|}{sector:  (-1,1)} \\ \hline
24 & $(1, 2, 1)_{0}$ & $18$ & $4$ & $-2$ & $-2$ & $0$ & $0$ & $-2$
& $2$ \\
%
27 & $(1, 1, 1)_{0}$ & $0$ & $10$ & $4$ & $-2$ & $0$ & $0$ & $-2$
& $2$ \\
%
%
29 & $(1, 1, 1)_{0}$ & $0$ & $-5$ & $1$ & $-5$ & $-3$ & $0$ & $-2$
& $2$ \\
%
%
31 & $(1, 2, 1)_{0}$ & $-18$ & $1$ & $1$ & $1$ & $-3$ & $0$ & $-2$
& $2$ \\
%
%
%
\hline \multicolumn{10}{|c|}{sector: (0,-1)} \\ \hline
34 & $(1, 1, 1)_{0}$ & $12$ & $-4$ & $4$ & $-4$ & $0$ & $-2$ & $2$
& $0$ \\
35 & $(1, 2, 1)_{0}$ & $-6$ & $2$ & $-2$ & $-4$ & $0$ & $-2$ & $2$
& $0$ \\
36 & $(1, 2, 1)_{0}$ & $-6$ & $2$ & $4$ & $2$ & $0$ & $-2$ & $2$ &
$0$ \\
%
%
38 & $(1, 1, 1)_{0}$ & $-24$ & $-7$ & $1$ & $-1$ & $3$ & $-2$ &
$2$ & $0$ \\
%
%
%
%
%
%
\hline \multicolumn{10}{|c|}{sector: (0,0)} \\ \hline
%
%
%
46 & $(1, 1, 1)_{0}$ & $12$ & $-4$ & $4$ & $0$ & $0$ & $2$ & $2$ &
$4$ \\
47 & $(1, 2, 1)_{0}$ & $-6$ & $2$ & $-2$ & $0$ & $0$ & $2$ & $-4$
& $-2$ \\
48 & $(1, 2, 1)_{0}$ & $-6$ & $2$ & $-2$ & $0$ & $0$ & $-4$ & $2$
& $-2$ \\
49 & $(1, 2, 1)_{0}$ & $-6$ & $2$ & $-2$ & $0$ & $0$ & $2$ & $2$ &
$4$ \\
\hline \multicolumn{10}{|c|}{sector: (0,1)} \\ \hline
50 & $(1, 1, 1)_{0}$ & $12$ & $-4$ & $4$ & $4$ & $0$ & $0$ & $2$ &
$2$ \\
51 & $(1, 2, 1)_{0}$ & $-6$ & $2$ & $-2$ & $4$ & $0$ & $0$ & $2$ &
$2$ \\
52 & $(1, 2, 1)_{0}$ & $-6$ & $2$ & $4$ & $-2$ & $0$ & $0$ & $2$ &
$2$ \\
%
%
%
%
56 & $(3, 1, 1)_{0}$ & $0$ & $5$ & $1$ & $1$ & $-3$ & $0$ & $2$ &
$2$ \\
%
%
%
%
\hline \multicolumn{10}{|c|}{sector: (1,-1)} \\ \hline
60 & $(1, 1, 1)_{0}$ & $-12$ & $-6$ & $-2$ & $-4$ & $0$ & $4$ &
$0$ & $0$ \\
61 & $(1, 1, 1)_{0}$ & $-12$ & $-6$ & $4$ & $2$ & $0$ & $4$ & $0$
& $0$ \\
%
%
%
64 & $(1, 2, 1)_{0}$ & $6$ & $3$ & $1$ & $-1$ & $-3$ & $4$ & $0$ &
$0$ \\
%
\hline \multicolumn{10}{|c|}{sector:  (1,0)} \\ \hline
%
%
%
%
%
70 & $(1, 2, 1)_{0}$ & $6$ & $3$ & $1$ & $-3$ & $3$ & $2$ & $0$ &
$-2$ \\
71 & $(1, 2, 1)_{0}$ & $6$ & $3$ & $1$ & $3$ & $-3$ & $2$ & $0$ &
$-2$ \\
72 & $(1, 1, 1)_{0}$ & $-12$ & $9$ & $1$ & $-3$ & $-3$ & $2$ & $0$
& $-2$ \\
%
%
74 & $(1, 1, 1)_{1}$ & $-12$ & $-6$ & $-2$ & $0$ & $0$ & $2$ & $0$
& $-2$ \\
%
%
\hline \multicolumn{10}{|c|}{sector: (1,1)} \\ \hline
%
%
77 & $(1, 1, 1)_{0}$ & $-12$ & $-6$ & $4$ & $-2$ & $0$ & $0$ & $0$
& $-4$ \\
%
%
%
80 & $(1, 2, 1)_{0}$ & $6$ & $3$ & $1$ & $1$ & $3$ & $0$ & $0$ &
$-4$ \\
81 & $(1, 1, 1)_{0}$ & $-12$ & $9$ & $1$ & $1$ & $-3$ & $0$ & $0$
& $-4$ \\ \hline
\end{tabular}
\end{table}

\mys{Violations of F-flatness} \label{violations}
In $N=1$ supersymmetric models such as the ones studied here, it
is generally the case that only isolated minima of the scalar
potential exist once all superpotential couplings, to all orders,
are taken into account. For theories with a large number of
fields, such as we study here, the number of isolated minima is
typically quite vast. Most of these minima are highly nontrivial,
involving cancellations between terms appearing in the
$\order(150)$ vanishing F-term conditions. Thus the ``flat
directions'' that we study here are most likely only approximate;
{\em i.e.} they are violated by some very high order terms in the
superpotential. However, provided that the vevs are small relative
to the Planck scale (which we assume) the shift in the vacuum away
from our approximately flat directions should be negligibly small.

For example, for the flat direction~(\ref{emm2}) in Model~1.2
studied in Section~\ref{case11}, it is easy to find superpotential
terms will violate F-flatness for this model. Suppose a monomial
$m$ of each of the fields involved in~(\ref{emm2}) with powers
$p_3$, $p_8$, etc.
\begin{equation}
m = S_{3}^{p_{3}}S_{8}^{p_{8}}S_{21}^{p_{21}}
S_{22}^{p_{22}}S_{29}^{p_{29}}S_{46}^{p_{46}}S_{72}^{p_{72}} .
\label{zzz} \end{equation}
Contributions to F-terms may occur if the $U(1)^8$ charge of~$m$
is zero or equal to the charge of any of the fields in the
spectrum. The assumption of vanishing charges implies
\begin{equation}
(p_3, \dots, p_{72}) = (a+b, b, a, b, b, a, a)
\end{equation}
where~$a$, $b$~are integers. Taking into account the string
selection rules for~$m$, we find they are only satisfied if and
only if $a + b = 0 \mod 3$, which implies a variety of degree~12
operators. Thus the violations of F-flatness from terms of the
form~(\ref{zzz}) first occur at degree~12. Taking $\lang S\rang
\sim 0.1$, we obtain F-term breaking of order $10^{-11} M_{\PL} =
10^7 \GeV$. This may seem large, but requires only a 1~part
in~$10^{10}$ shift in the vevs to cancel, moving to a true minimum
of the scalar potential. This should not change the dominant
features of the effective low-energy mass matrices and Yukawa
couplings.

A similar (tedious) analysis could be carried out for all proposed
flat directions, for all models, identifying the lowest order at
which nonvanishing contributions to flat directions would occur.
However, since we have already verified stringent F-flatness to
degree~9, the lowest order we will ever find is degree~10.
Generalizing the arguments made above, this would merely require a
1~part in~$10^{9}$ shift in the vevs, which again has negligible
effects in the low-energy theory. For this reason we believe that
the high orders to which we have checked F-flatness should suffice
for our purposes.

Admittedly then, the flat directions we study are only a tiny
sample of the vast number of approximate minima. Nevertheless,
since they are the easiest to classify and do not require detailed
knowledge of the strengths of superpotential couplings, they are
the most sensible class of approximate minima to study in a first
detailed analysis of the low-energy couplings.

\mys{Selection rules for superpotential couplings} \label{selapp}
In this appendix, we review constraints on superpotential
couplings in $Z_3$ constructions such as the BSL$_A$ models that
we study.  Orbifold selection rules are presented from the
practical standpoint: we explain how they are implemented rather
than why they are true.  The origin of these rules in the
underlying conformal field theory has been described in detail
in~\cite{Font:1988tp} and reviewed in Appendix~B
of~\cite{Font:1989aj}. The presentation here rests on
Refs.~\cite{Friedan:1985ge,Hamidi:1986vh,Dixon:1986qv,Dixon:1987bg}.

\bfe{Gauge invariance.} This is very restrictive in the BSL$_A$
models, where the gauge groups are rank 16.  The $U(1)$ parts of
the gauge group, $U(1)^8$ or $U(1)^9$ factors, greatly reduce the
number of invariants beyond what would be allowed from non-abelian
factors alone. This is particularly true because of the large
number of matter fields that are non-Abelian singlets.  These
matter fields are never singlets with respect to all of the $U(1)$
factors.

\bfe{Point group selection rule.} This is a {\it twisted triality}
invariance.  Non-oscillator twisted fields $T$ and oscillator
twisted fields $Y$ ({\it blowing up modes} of the (0,2)
construction~\cite{Font:1988tp}) have twisted triality 1, whereas
untwisted matter fields $U$ and K\"ahler moduli $M^{k \bar \ell}$
have twisted triality 0. The point group selection rule for a
$Z_3$ model states that only couplings of the form $M^\ell U^m
T^{3n} Y^{3p}$, where $\ell,m,n,p \in \Zbf$, are allowed.  These
are just couplings of vanishing twisted triality; i.e., triality
of $0 \mod 3$.

\bfe{Lattice group selection rule.} This is a restriction on
couplings between twisted sector fields, and is a 3-fold triality.
Each twisted matter field has fixed point labels $(n_1,n_2,n_3)$,
with each entry taking values $0,\pm 1$.  An invariant coupling
must have vanishing triality with respect to each of the entries.
Thus, consider a coupling with $m$ twisted fields \beq
T_{n_1^{(1)},n_2^{(1)},n_3^{(1)}} \cdots
T_{n_1^{(m)},n_2^{(m)},n_3^{(m)}} . \eeq The lattice group
selection rule requires \beq n_i^{(1)} + \cdots + n_i^{(m)} = 0
\mod 3, \quad \forall \; i=1,2,3. \eeq

\bfe{H-momentum conservation.}  This takes its name from the
bosonized description of Neveu-Schwarz/Ramond world-sheet
fermions, $:\psi^{2m-1}  \psi^{2m}: (z) \equiv \p H^m(z)$. Here,
$z$ is the complex worldsheet coordinate of the underlying
conformal field theory. The intricacies of this selection rule
have been reviewed, for example, in~\cite{Font:1989aj}. Here we
merely state the results as they pertain to superpotential
couplings.\footnote{While this selection rule is most often
described in terms of the covariantly quantized
string~\cite{Friedan:1985ge}, it is also straightforward to derive
in a lightcone description of the physical states.} A general
n-point amplitude associated with the superpotential vertex \beq
\int d^2 \theta \; W \ni \chi_1 \chi_2 \phi_3 \ldots \phi_n \eeq
in the effective supergravity arises from a correlation function
in the underlying conformal field theory of the form \beq \langle
\, V_{-\half}(z_1) \, V_{-\half}(z_2) \, V_{-1}(z_3) \, V_0(z_4)
\, \cdots \, V_0 (z_n) \, \rangle \eeq Here, the subscripts
indicate ghost number $q$ of vertex operators in the covariant
formulation~\cite{Friedan:1985ge}.

For the untwisted matter states, what is important is that they
have a degeneracy of 3 corresponding to different internal
$SO(2)^3$ weights.  Each $SO(2)$ factor corresponds to one of the
three complex planes of the compact space.  The states carry a
label of the degeneracy: $U^i$ where $i=1,2,3$.  The different
$SO(2)^3$ weights determine different combinations of $H(z)$ that
appear in the vertex operators.  It is this $H(z)$ dependence that
is important to H-momentum conservation.  Thus constraints arise
on the labels $i$ that can appear in an invariant coupling.  It
will later prove important that the operators with ghost number
$q=0$ contain a worldsheet derivative factor \beq V_0(U^i)  \sim
\p X^i \label{untwvert} \eeq whereas the others do not. Here,
$X^i$ are the (complex) worldsheet bosons corresponding to the 6D
compact space.  Worldsheet derivatives such as $\p X^i$ are
important in the final selection rule discussed below.

For the twisted sector there is no such degeneracy of $SO(2)^3$
weights.  Instead the degeneracy corresponds to the fixed point
labels discussed above.  These play no role in H-momentum
conservation.  What is important is that for non-oscillator
twisted fields \beq V_0(T) \sim \sum_i f^i(H) \pX^i \label{twvert}
\eeq where $f^i(H)$ stands for details that we will not discuss
here, but which are involved in H-momentum conservation. In the
case of oscillator twisted fields $Y^\ell$, all their vertex
operators pick up an extra derivative factor, corresponding to
left-moving oscillator number $N_L=1/3$ \beq V_q(Y^\ell) \sim
V_q(T) \pXb^\ell . \label{oscvert} \eeq

The 9 K\"ahler moduli $M^{k \bar \ell}$ also play a role in the
superpotential couplings.  Since we are only interested in how
they interact with matter, we can always include them through
$V_0$ operators, which take the form \beq V_0(M^{k \bar \ell})
\sim \pX^k \pXb^{\bar \ell} . \eeq

A trilinear coupling between untwisted matter fields, $U^i U^j
U^k$, only conserves H-momentum if $i$, $j$ and $k$ are different
from each other.  Thus \beq U^i U^j U^k \sim \e^{ijk}. \eeq For a
higher order coupling, the $V_0$ operators are not constrained by
H-momentum \beq U^i U^j U^k U^{\ell_1} \cdots U^{\ell_{n}} \sim
\e^{ijk} \pX^{\ell_1} \cdots \pX^{\ell_n} . \label{notw} \eeq With
respect to the degeneracy labels on the untwisted fields, which
serve as generation labels in the effective supergravity, many
different couplings are allowed, corresponding to different
assignments of $n$ fields to the $q=0$ picture.

A trilinear coupling between twisted matter fields always
conserves H-momentum. For a higher order twisted coupling,
H-momentum conservation constraints on the $f^i(H)$ in
\myref{twvert} picks out a certain combination of the derivatives:
\beq T^{3m} \sim (\p X^1 \p X^2 \p X^3)^{m-1} . \label{eqn005}
\eeq  If there are both twisted and untwisted fields, then it is
convenient to take the untwisted fields in the $q=0$ picture, so
that \beq T^{3m} U^{i_1} \cdots U^{i_n} \sim (\p X^1 \p X^2 \p
X^3)^{m-1} \pX^{i_1} \cdots \pX^{i_n} . \label{lser} \eeq

It is obvious how the above expressions are modified once
oscillator twisted states or K\"ahler moduli are included. For
example,
\begin{eqnarray}
T^{3m-1} Y^\ell &\sim& (\p X^1 \p X^2 \p X^3)^{m-1} \pXb^\ell,
\nonumber \\
U^i U^j U^k M^{m \bar \ell} &\sim& \e^{ijk} \pX^m \pXb^{\bar \ell}
.
\end{eqnarray}

\bfe{Automorphism selection rule.} This last selection rule
requires that couplings be invariant under automorphisms of the
$SU(3)^3$ lattice. Here this amounts to examining the factors of
$\p X^i$ and $\pXb^i$ coming from the vertex operators.  Suppose
\begin{widetext}
\begin{equation}
\langle \, V_{-\half}(z_1) \, V_{-\half}(z_2) \, V_{-1}(z_3) \,
V_0(z_4) \, \cdots \, V_0 (z_n) \, \rangle \sim \prod_{i=1}^3
(\pX^i)^{m_i} (\pXb^i)^{p_i} .
\end{equation}
\end{widetext}
Then the automorphism selection rule states that the coupling will
vanish unless \beq m_i - p_i = 0 \mod 3 \quad \forall \quad i .
\label{autorule} \eeq

It is convenient to define powers that do not distinguish between
the indices, just counting the number of $\pX$s or $\pXb$s that
appear: \beq m=\sum_i m_i, \qquad p=\sum_i p_i . \eeq A necessary
but not sufficient condition is that \beq m-p=0 . \mod 3
\label{yeriw} \eeq Twisted fields contribute to $m-p$, $\mod 3$,
only if they are oscillators, through \myref{oscvert}. This is
because \myref{eqn005} only gives multiples of 3.  Each untwisted
oscillator subtracts 1 from $m-p$, $\mod 3$. Untwisted superfields
contribute to $m-p$ through \myref{untwvert}.  In couplings
without twisted fields \myref{notw}, $m-p$ just counts the number
of untwisted fields $\mod 3$. In couplings with twisted fields
\myref{lser}, we can always associate the untwisted fields with
$V_0$ operators, so again $m-p$ counts the number of untwisted
fields $\mod 3$. Finally, the K\"ahler moduli do not contribute to
$m-p$, $\mod 3$. From these considerations it can be seen that a
convenient way to encode the constraint \myref{yeriw} is the
following: we assign {\it untwisted triality} of $1$ to fields $U$
and $-1$ to fields $Y$.  Fields $T$ and $M^{k \bar \ell}$ have
untwisted triality $0$. The demand that couplings be invariant
with respect to untwisted triality is equivalent to \myref{yeriw}.

It is easy to show that if a coupling has vanishing untwisted
triality, so that it satisfies \myref{yeriw}, then it can be made
to satisfy the stricter automorphism selection rule
\myref{autorule} simply by supplementing with an appropriate
combination of off-diagonal moduli. As an example, let us examine
couplings of the form $TTTTTT Y^1 Y^1 Y^2$, with none of the $T$
fields an oscillator state.  This involves nine twisted states, so
remembering the derivative terms which come from $Y^i$ and using
(\ref{eqn005}) we find
\begin{widetext}
\begin{equation}
TTTTTT Y^1 Y^1 Y^2 \sim (\p X^1 \p X^2 \p X^3)^2 \bar \p \bar X^1
\bar \p \bar X^1 \bar \p \bar X^2.
\end{equation}
\end{widetext}
Now apply the rule \myref{autorule}: \beq m_1-p_1=0, \quad
m_2-p_2=1,  \quad m_3-p_3=2. \eeq Thus, the coupling is forbidden
by the automorphism selection rule.  However, notice that \beq
M^{3 \bar 2} \sim \p X^3 \bar \p \bar X^2 \eeq provides just the
factors we need in order to satisfy the automorphism selection
rule. Consequently, the coupling \beq M^{3 \bar 2} TTTTTT Y^1 Y^1
Y^2 \eeq is allowed.

More generally, suppose
\begin{equation}
m_i - p_i = 3 \ell_i + r_i, \qquad r_i \ni \{ -1, 0, 1 \} .
\end{equation} Thus if the coupling has vanishing untwisted triality,
$r_1+r_2+r_3 = 0 \mod 3$.  Then it is easy to check that for any
choice of the $r_i$ it is possible to find a combination of
off-diagonal moduli that will cancel the $r_i$'s. For example if
$(r_1,r_2,r_3)=(-1,0,1)$, then $M^{1 \bar 3}$ will do the job. Or,
if $(r_1,r_2,r_3)=(1,1,1)$ then $M^{3 \bar 1} M^{3 \bar 2}$ will
suffice.  The other nontrivial possibilities are just permutations
of $(-1,0,1)$, or $(-1,-1,-1)$.  It is easy to check that these
can be compensated in a manner similar to the two examples just
given.

The off-diagonal moduli parameterize angles between the three
complex planes of the compact space.  If we were to assume that
the vacuum was maximally symmetric, then the vevs of the
off-diagonal moduli would vanish. We will not make this assumption
here, since it is a very special point in moduli space and hardly
corresponds to a generic situation.  Since we allow for
nonvanishing off-diagonal moduli, any coupling that satisfies
\myref{yeriw} but not \myref{autorule} can be made to satisfy
\myref{autorule} simply by adding some number of off-diagonal
moduli.  Thus in the presence of off-diagonal moduli, we need only
check untwisted triality to ensure that both H-momentum
conservation and the automorphism selection rules are satisfied.

A simple example is afforded by degree~5 superpotential couplings.
A coupling of all untwisted fields will not work because the
untwisted triality is $5 \simeq 2$.  Given that twisted fields
must be included, twisted triality requires exactly~3.  But then 2
untwisted remain, which give untwisted triality of 2.  To cancel
this, 2 of the twisted fields must be oscillators.  Thus the
unique type of trialities-allowed coupling is $UUYYT$.  In our
analysis of the BSL$_A$ models, we find that this is never gauge
invariant, and thus never allowed. A further simple consequence of
untwisted triality is that models that do not contain twisted
oscillators (all those that fall into Patterns~2.6, 4.5, 4.7
and~4.8) can only have superpotential couplings whose degree is a
multiple of~3. This is seen explicitly in Table~\ref{spcpn}.

\bfe{Summary.} Provided that we do not go to a special point in
the moduli space where off-diagonal moduli vanish, the imposition
of the selection rules just amounts to the simultaneous
satisfaction of: (i) lattice triality, (ii) twisted triality,
(iii) untwisted triality. This makes the selection rules very easy
to automate and has greatly aided our analysis.


\end{document}